\documentclass[twocolumn,linenumbers,tighten]{aastex631}

\usepackage[caption=false]{subfig}
\usepackage{appendix}
\newcommand{\gskfont}{
  \color{red}
}

\DeclareTextFontCommand{\gsk}{\gskfont}

\shorttitle{Formation of The Lyman Continuum During Solar Flares}
\shortauthors{McLaughlin et al.}
\graphicspath{{./}{figures/}}

\begin{document}

\title{Formation Of The Lyman Continuum During Solar Flares} % \footnote{Released on April, 21st, 2022}}

\author[0000-0002-6900-0936]{Shaun A. McLaughlin}
\affiliation{Queen's University Belfast, University Rd, Belfast BT7 1NN, Northern Ireland}

\author[0000-0001-5031-1892]{Ryan O. Milligan}
\affiliation{Queen's University Belfast, University Rd, Belfast BT7 1NN, Northern Ireland}
\affiliation{Department of Physics Catholic University of America, 620 Michigan Avenue, Northeast, Washington, DC 20064, USA}

\author[0000-0001-5316-914X]{Graham S. Kerr}
\affiliation{Department of Physics Catholic University of America, 620 Michigan Avenue, Northeast, Washington, DC 20064, USA}
\affiliation{Solar Physics Laboratory (Code 671), Heliophysics Science Division, NASA Goddard Space Flight Center, Greenbelt, MD 20771, USA}

\author[0000-0002-3305-748X]{Aaron J. Monson}
\affiliation{Queen's University Belfast, University Rd, Belfast BT7 1NN, Northern Ireland}

\author[0000-0002-4819-1884]{Paulo J. A. Sim\~{o}es}
\affiliation{Center for Radio Astronomy and Astrophysics Mackenzie, School of Engineering, Mackenzie Presbyterian University, S\~{a}o Paulo, Brazil}
\affiliation{SUPA, School of Physics \& Astronomy, University of Glasgow, Glasgow G12 8QQ, UK}

\author[0000-0002-7725-6296]{Mihalis Mathioudakis}
\affiliation{Queen's University Belfast, University Rd, Belfast BT7 1NN, Northern Ireland}

%\collaboration{6}{(AAS Journals Data Editors)}

\begin{abstract}
%\vspace{10cm}
The Lyman Continuum (LyC; $<911.12$\AA) forms at the top of the chromosphere in the quiet-Sun, making LyC a powerful tool for probing the chromospheric plasma during solar flares. To understand the effects of non-thermal energy deposition in the chromosphere during flares, we analysed LyC profiles from a grid of field-aligned radiative hydrodynamic models generated  using the RADYN code as part of the F-CHROMA project. The spectral response of LyC, the temporal evolution of the departure coefficient of hydrogen, $b_1$, and the color temperature, $T_c$, in response to a range of non-thermal electron distribution functions, were investigated. The LyC intensity was seen to increase by 4-5.5 orders of magnitude during solar flares, responding most strongly to the non-thermal electron flux of the beam. Generally, $b_1$ decreased from $10^2$--$10^3$ to closer to unity during solar flares, indicating a stronger coupling to local conditions, while $T_c$ increased from $8$--$9$~kK to $10$--$16$~kK. $T_c$ was found to be approximately equal to the electron temperature of the plasma when $b_1$ was at a minimum. Both optically thick and optically thin components of LyC were found in agreement with the interpretation of recent observations. The optically thick layer forms deeper in the chromosphere during a flare compared to quiescent periods, whereas the optically thin layers form at higher altitudes due to chromospheric evaporation, in low-temperature, high-density regions propagating upwards. We put these results in the context of current and future missions.

\end{abstract}

\section{Introduction} 

Solar flares are the most energetic events to occur in the solar system, releasing up to $10^{32}$ erg of energy over the course of tens of minutes \citep{Fletcher_2011}. This energy is typically thought to be released via magnetic reconnection in the solar corona. Following release, this energy manifests itself in several ways, including in the acceleration of large quantities of charged particles that propagate along magnetic field lines, deeper into the Sun's atmosphere. These charged particles undergo Coulomb collisions with the ambient plasma depositing energy in the solar chromosphere, where intense heating, ionization, and mass flows result. The subsequent increase in radiation allows us to shed light on the physical mechanisms that release and transport energy. Modelling by \citet{Allred_2005} suggests that solar flare emission is energetically dominated by the Balmer continuum. However, Balmer continuum observations have been severely lacking. Lyman Continuum flare observations have been more readily available in recent years from the Extreme Ultraviolet Experiment (EVE; \citealt{Woods2012}) on board the Solar Dynamic Observatory (SDO; \citealt{Pesnell2012}), with the potential of upcoming flare observations from the Spectral Imaging of the Coronal Environment (SPICE; \citealt{Spice2020}) instrument on board the Solar Orbiter \citep{Solarorbiter2020}, and the EUV High-Throughput Spectroscopic Telescope (EUVST; \citealt{EUVST}) on board Solar-C \citep{SOLARC2014}. 

The Lyman Continuum (LyC; $<$911.12\AA) results from the free-bound transition of a free electron to the ground state of an ambient hydrogen nuclei. In the quiet-Sun, LyC forms at the top of the chromosphere/base of the transition region where the temperature rises from the cooler, denser chromosphere ($T\leq10^{4}$K) to the corona, ($T\geq10^{6}$K; \citealt{Vernazza1973,Vernazza1976,Vernazza1981,Averett2008,Machado2018}). Therefore, the LyC is sensitive to energy perturbations in the chromosphere induced during solar flares,  and since thermalization occurs very rapidly at the higher densities located here, its spectrum may reflect the local plasma temperature \citep{Noyes1970}. The LyC is a potentially powerful diagnostic tool of the chromospheric response to flare energy injection, but this potential is presently largely untapped.  

%LyC spectral radiance can be approximated by an Eddington-Barbier relation, see Section~\ref{Maths} \citep{Noyes1970,Vernazza1972}. Using the Eddington-Barbier approximation to fit the LyC spectrum, values of the departure coefficient of the first level of hydrogen, $b_1$, and the color temperature, $T_c$, can be determined. If $b_1$ is close to unity, then $T_c$ is approximately equal to the electron temperature of the plasma, $T_e$. The thermodynamic state of the chromospheric plasma can therefore be determined solely from LyC spectra, independent of any modelling assumptions. 

One of the earliest investigations into LyC formation during solar flares was performed by \citet{Machado1978}, who observed nine flares using the extreme-ultraviolet (EUV)  spectrograph on the Apollo Telescope Mount onboard Skylab \citep{Apollotelescope1974}. They found that the non-local thermodynamic equilibrium (NLTE) departure coefficient of the first level of hydrogen ($b_1$) tended towards unity during flares, more so than in active regions or the quiet-Sun. The departure coefficient is given by $n_1/n_1^*$, where $n_1^*$ is the ground state LTE population and $n_{1}$ is the NLTE ground state populations, meaning the LyC formation region was driven closer to LTE conditions during the solar flares  \citep{Menzel1937}. They also found that the color temperature ($T_c$) derived from the head of the continuum was around $T_{c}\sim7.9$--$9$~kK, comparable to that of the surrounding quiet-Sun, but hotter than active regions. They concluded that in flares LyC is optically thick and forms close to LTE in a region of higher density deeper in the chromosphere, whereas in active regions or the quiet-Sun $b_1\approx10^2$ \citep{Noyes1970}. However, they also showed that $T_c$ measured at shorter wavelengths ($700$--$790$~\AA) revealed higher temperatures ($10$--$15$~kK, although with large uncertainties), suggesting the presence of an optically thin LyC layer that formed higher in the atmosphere. One limitation of this study was that several of the observations were carried out in spectral scanning mode; spectra were obtained in a 5~arcsec$^2$ area over $300$--$1335$~\AA, over 3.8 minutes, meaning that different parts of the spectrum may have been sampled at different times and may not reflect true changes due to the flare.

%They found that the LyC intensities were greatly enhanced during solar flares, increasing by several orders of magnitude compared to the quiet-Sun intensities. Applying the Eddington-Barbier relation to the observed LyC spectra, they found $b_1$ was close to unity during solar flares and determined $T_c$ to be approximately $7900$--$9000$K. They stated this is evidence for an optically thick component of LyC, formed deeper in the chromosphere close to LTE during solar flares. Lastly, \citet{Machado1978} found that $b_1$ and $T_c$ values determined at shorter wavelengths ($\lambda\leq790$\AA) were greater than those found nearer the head of the continuum. They hypothesized that the increase in $T_c$ at shorter wavelengths was evidence for an optically thin component of LyC formed higher up in the chromosphere during solar flares. One limitation of this study was that several of the observations were carried out in spectral scanning mode meaning that different parts of the spectrum may have been sampled at different times and therefore may not reflect true changes due to the flare.

More recently, \citet{Machado2018} performed a study using Sun-as-a-star observations from EVE on board SDO which allowed the temporal evolution of LyC emission during six major solar flares to be investigated. After converting the EVE data from spectral irradiance to specific intensity, by assuming flaring areas, they reported that the LyC intensities at the head of the continuum were enhanced by 3--4 orders of magnitude relative to preflare values. Their results supported the earlier conclusions of \cite{Machado1978}. However, \citet{Machado2018} found a larger color temperature, noting that $T_{c}$ determined between $870$--$912$~\AA\ increased from $T_{c} = 8$--$9$~kK to $T_{c}\approx9$--$12$~kK, perhaps due to the study of more energetic events.

\cite{Milligan2012} reported an enhancement in LyC emission during an X-class flare that occurred on 2011 February 15 also using data from SDO/EVE. A follow-up study showed that the total energy radiated by LyC was a few percent of the total non-thermal electron energy, amounting to $1.8$$\times$$10^{29}$~erg over the course of the flare, comparable to the radiated soft X-ray energy \citep{Milligan2014}. LyC irradiance increased by approximately a factor of 10 during the impulsive phase. 
%Despite the Lyman continuum being a prominent radiator in the chromosphere during solar flares, the literature has been sparse over the past 20 years. 

%They found $b_1$ decreased from $10^{2-3}$ in the QS to approximately unity during solar flares, in agreement with the literature \citep{Machado1978,Ding1997,Lemarie_2004}. They determined $T_c$ to increase from approximately $8000$K to $10000$--$12000$K, much higher than that observed by \citet{Machado1978}, but they have attributed the difference arising due to the more energetic flares analysed. Also, \citet{Machado2018} only observed one event that showed an increase in $T_c$ at shorter wavelengths, unlike \citet{Machado1978}, who observed this in all nine events.

\citet{Lemarie_2004} presented observations of an X5.3 solar flare from the head of the LyC obtained from scattered light detected by the Solar Ultraviolet Measurements of Emitted Radiation spectrometer (SUMER; \citealt{1995Sumer}) on board the Solar \& Heliospheric Observatory (SOHO; \citealt{1995SOHO}) mission. They determined the local increase of LyC radiance to be a factor of several thousand, in agreement with \citet{Machado1978}. Using the ratio of the spectral radiance at 910\AA\ and 890\AA\ they estimated $T_c = 12.15$~kK. %LyC observations from SOHO/SUMER have also been utilized in the study of solar prominences; \citet{Parenti2005} found that two areas of a prominence had {\it lower} $T_{c}$ than the quiet-Sun value. GSK: I dont think we need the prominence discussion
%$b_1$=48 and %I don't think you can estimate b1 from flux ratios, just T

%The LyC has not only been utilized during solar flares but also during solar prominences, in active regions, and during QS observations  \citep{Hirayama1985,Parenti2005}. \citet{Parenti2005} analysed data during a solar prominence using the SUMER spectrometer on board SOHO. They state that as prominences are highly complex structures, it is important to understand the mechanisms in the different layers. As $T_c$ is reflective of the electron temperature for kinetic temperatures below 15000K, the LyC is a powerful tool for understanding solar prominences \citep{Gouttebroze1993}. \citet{Parenti2005} found that two areas of the prominence had different $T_c$/$T_e$ temperatures despite having similar brightness temperatures. Area 1 had a $T_c$=8281K, and Area 2 had a $T_c$=7564K. They note that this is due to the differing properties of the two prominence areas. Area 1 appeared to be more inhomogeneous, with a large transition region and coronal contributions. Both areas had a lower $T_c$ than the QS value, $T_c$= 8753K.

There have been several attempts at modelling LyC during solar flares. \citet{Ding1997} calculated LyC intensities in response to a precipitating beam of non-thermal particles using a non-LTE, radiative transfer code. They found that LyC is very sensitive to the incoming flux of electrons, and they predicted that the temperature at the formation height of LyC will increase in conjunction with a downward shift of the transition region. In particular, they noted that non-thermal collisions were responsible for driving down $b_1$. In their experiments, \citet{Ding1997} manually modified semi-empirical model atmospheres to investigate the impact of the chromospheric temperature gradients, and transition region depth on the LyC formation, in a manner that was not self-consistent (i.e. they did not simulate the time-dependent chromospheric radiation-hydrodynamic response to energy injection by a beam of non-thermal electrons). For the purposes of including non-thermal collisional ionization of hydrogen, they assumed non-thermal electron fluxes of $F = [10^{10}, 10^{11}, 10^{12}]$~erg~cm$^{-2}$~s$^{-1}$, with spectral indices $\delta = [3, 4, 5]$, and a fixed low-energy cutoff $E_c = 20$~keV. They found that LyC intensities increased by $1$--$3$ orders of magnitude compared to quiet-Sun values, and that the LyC intensities were highly sensitive to the flux of the non-thermal electron beam. The LyC intensities were less sensitive to the spectral index of the non-thermal electron distribution.

%LyC has been proposed as a diagnostic of non-thermal process during solar flares by \citet{Ding1997} who modelled the spectral response to an increase in the chromospheric plasma temperature, a downward shift in the transition region, and a precipitating non-thermal electron beam. 
%Again applying the Eddington-Barbier approximation, they determined $b_1$ and $T_c$ values for the spectra, the values of which are in agreement with the literature discussed above. However, they did not investigate the wavelength dependence of $b_1$ and $T_c$. 
% using their iterative NLTE radiative hydrostatic code

More recent modelling of LyC emission was performed by \citet{Druett2019} using the radiation transport and hydrodynamic code, HYDRO2GEN. They investigated the spectral response of LyC to various electron beam fluxes during the impulsive and gradual phases of solar flares. They found that the LyC intensities increased by 2 orders of magnitude from their F10 model to their F12 model, highlighting the sensitivity of the LyC intensities to the flux of the non-thermal electron beam. From continuum contribution functions, they determined that LyC forms under optically thick conditions at the top of the chromosphere. During the impulsive phase, they noted a flattening of the LyC spectra but found that the gradient steepened further from the head of the continuum, in agreement with \citet{Machado1978} and \citet{Machado2018}. During the gradual phase, however, they observed a flattening of the LyC spectra at shorter wavelengths compared to the impulsive phase spectra. They stated that the flattening below 700\AA\ is due to emission from high-altitude regions that cool sufficiently to allow recombinations to occur, resulting in an enhanced intensity longward of $\sim700$~\AA.

In this paper, we build upon these prior modelling efforts, by investigating the plasma properties (namely $T_{c}$ and $b_{1}$) in the region of LyC formation, and the diagnostic potential of the LyC in flares. We present a detailed analysis of LyC emission from a large parameter space of solar flare models using the Flare CHRomospheres: Observations, Models and Archives (F-CHROMA\footnote{https://star.pst.qub.ac.uk/wiki/public/solarmodels/start.html}) grid of models generated by the 1D field-aligned radiative-hydrodynamic code RADYN. %Section~\ref{METHOD} describes the RADYN code, the F-CHROMA grid of models, and the analysis techniques used. The results are presented in Section~\ref{RESULTS}, with a discussion of these provided in Section~\ref{DISCUSSION}. The conclusions of our findings are summarised in Section~\ref{CONCLUSION}. 

\section{Numerical Simulations}\label{METHOD}

RADYN is a 1D field-aligned radiative hydrodynamic code that solves the coupled time-dependent equations of hydrodynamics, NLTE radiative transfer, non-equilibrium atomic level populations, and charge conservation \citep{Carlson1992,Carlson1995,Carlson1997,Allred_2005,Allred2015}. The code's adaptive grid \citep{Dolf1987} allows RADYN to resolve steep gradients or shocks that easily form in flares. %In solar flare simulations, the weighting of this grid is typically used to resolve velocity and temperature features in the transition region \citep{Allred2006}.
RADYN solves the NLTE radiation transfer for a 6-level hydrogen atom, 9-level helium atom, and a 6-level \ion{Ca}{2} ion. 

Important for modelling the LyC, RADYN models the non-equilibrium populations, and for hydrogen and helium includes non-thermal collisional rates between  the injected non-thermal electron distribution and ambient plasma. For helium, we use the expressions from \cite{1985A&AS...60..425A} to model non-thermal collisional ionisation of \ion{He}{1}$\rightarrow$\ion{He}{2}, and from \ion{He}{2}$\rightarrow$\ion{He}{3}. For hydrogen, we include non-thermal collisional ionisation from the ground state, and excitation from the ground state to $n=[2,3,4]$, following the approach of \cite{Fang1993}. In this approach, we are able to include the important effects of secondary collisions. As discussed in several prior studies \citep[e.g.][]{1983ApJ...272..739R,Fang1993,2004A&A...416L..13K,2009A&A...499..923K}, non-thermal collisional ionisation from the excited states are of lesser importance, in part because the population of the excited states are orders of magnitude smaller than the ground state. Thus the ionisation stratification is unlikely to be drastically altered by the omission of non-thermal collisions from the ground state. Note, however, that some models have included non-thermal effects from the excited states \citep[see e.g.][]{1993SoPh..143..259Z,2018A&A...610A..68D} though it is unclear just how much of a role those collisions played compared to those from the ground state. Figure 3 of \cite{1993SoPh..143..259Z} does show that non-thermal collisions from $n=2$ became important relative to the thermal collisional rates, but only deep in the atmosphere. For the reasons that we discussed above, and since we are focused on the LyC which forms in the upper chromosphere/lower transition region, we believe that we are justified in using only ground state non-thermal collisional rates including secondary effects. %For full model details consult \cite{Allred2015} and references below. %%GSK: I don't see why mentioning Mg II here is necessary so just deleted all that section... really, if we include a model atom we can include whatever species we want so no point in mentioning what we don't include

RADYN is a well-established resource that has been extensively used to model the response of the solar atmosphere to flare energy injection \citep[e.g.][]{Abbet1999,Allred_2005,Allred2015, 2015SoPh..290.3487K, 2017ApJ...836...12K,2022ApJ...928..190K, 2018ApJ...852...61K, 2015A&A...578A..72K, 2016ApJ...827...38R, 2016ApJ...827..101K, 2019ApJ...871...23K, 2020ApJ...900...18K,2021ApJ...912..153K, 2017A&A...605A.125S,Brown2018, 2020ApJ...895....6G}. When simulating flares driven by electron beams (the `standard model') the non-thermal electron distribution is modelled as a power law characterised by the spectral index, $\delta$, the low-energy cutoff, $E_c$, and energy flux density, $F$. The transport and thermalization of these non-thermal electrons are modelled by solving the Fokker-Planck equations, including various diffusion terms \citep{Allred2015}. Note that we solve the Coulomb operator using general forms of the Rosenbluth potentials and so \textsl{no assumption is made as to the target temperature}; that is, there is no need to make a cold or warm target approximation \citep[see also][]{Allred_2020}. Beam parameters can be obtained through backward modelling of hard X-ray observations taken with space-borne satellites such as the Ramaty High Energy Solar Spectroscopic Imager (RHESSI; \citealt{RHESSI2002}) or the Fermi Gamma-ray Space Telescope \citep{Fermi2009}, usually under the assumption of thick target collisions \citep{Brown1971,Holman2003,Krucker2007,Krucker2008,Kontar2008}. In our study, we do not model a specific flare, but instead survey a large parameter space of $\delta$, $E_{c}$ \& $F$ values, with the values of those parameters guided by typical observed ranges \citep[e.g.][]{2011SSRv..159..301K,2011SSRv..159..107H}.

The F-CHROMA grid of RADYN flare models, hosted by Queen's University Belfast, consists of 72 RADYN simulations \cite[][]{FCHROMAinprep}. This grid includes $\delta = [3,4,5,6,7,8]$, and $E_{c} = [10,15,20,25]$~keV, and models various magnitudes of injected non-thermal electron flux as a $20$~s triangular profile in time, peaking at $t=10$~s. The time-integrated energy flux ranges $F_{tot} = [
3\times10^{10}, 1\times10^{11}, 3\times10^{11}, 1\times10^{12}]$~erg~cm$^{-2}$. It is typical in the flare community to refer to the energy flux density, and so for the remainder of this paper we quote the peak injected flux (at $t=10$~s in each simulation), which are $F = [3\times10^{9}, 1\times10^{10}, 3\times10^{10}, 1\times10^{11}]$~erg~cm$^{-2}$~s$^{-1}$. For brevity, we refer to these simulations as, e.g., $1F11$, equating to $1\times10^{11}$~erg~cm$^{-2}$~s$^{-1}$.

In this study, we have selected 25 of the 72 models (Table~\ref{Table:1}). Not every combination of $\delta$ and $E_{c}$ were modelled for inclusion in this grid for the highest energy flares. The models omitted from our subset are the 1F11, $E_c$=20keV, $\delta = [6,7]$, models and the 1F11, $\delta = 5$, $E_c$=15keV model.

\begin{table}[ht!] 
\centering
 \begin{tabular}{|c|c|c|} 
 \hline
  $F$ (erg$~$cm$^{-2}~$s$^{-1}$)& $\delta$ & $E_c$(kev)\\
 \hline
 \hline
 $3\times10^{9}$ & 3, 4, 5, 6, 7 & 15, 20, 25\\
 \hline
 $1\times10^{10}$& 3, 4, 5, 6, 7 & 15, 20, 25\\
 \hline
 $3\times10^{10}$& 3, 4, 5, 6, 7 & 15, 20, 25\\
 \hline
 $1\times10^{11}$& 3, 4, 5 & 20, 25\\
 
 \hline
\end{tabular}
\caption{Non-thermal electron beam parameters used for generating the RADYN models analysed in this work.}
\label{Table:1}
\end{table}

\section{LyC Response to Flare Energy Injection}

\begin{figure*}
\includegraphics[width=1\textwidth]{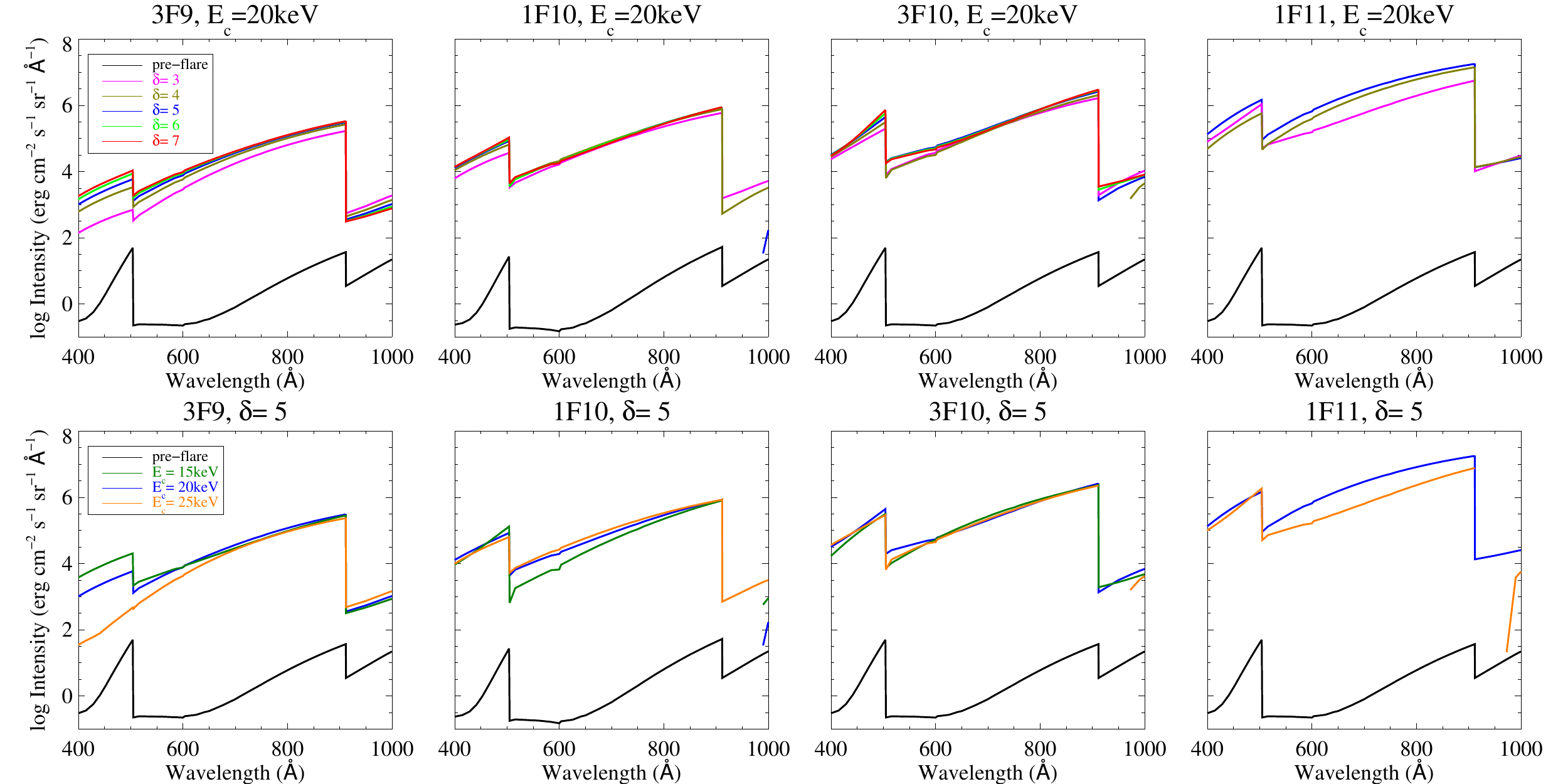}
\caption{Synthetic RADYN continua spectra from 400--1000\AA\ showing the head of the \ion{He}{1} continuum ($<$503.98\AA), LyC ($<$911.12\AA), and the tail of the \ion{Ca}{2} continuum ($<$1044.00\AA). The top row shows spectra for a fixed $E_{c} = 20$~keV, with peak beam fluxes of $3F9$, $1F10$, $3F10$, and $1F11$, all shown at the peak in the LyC spectrum (between 9.7~s--13.6~s for all models). The black curve denotes the preflare spectra, while the colored curves show the spectra for spectral indices of $\delta$=3--7. The bottom row shows the spectra but for a fixed $\delta$=5, varying $E_c$=15, 20, and 25 keV. Note the $3F10$, $\delta$=5, $E_{c} = 20$~keV and the $1F10$, $\delta$=5, $E_{c} = 20$~keV models have transient negative intensities in the tail of the LyC continuum at the time of the peak LyC emission. This is due to numerical noise in the simulation at these times, which only lasts a very short amount of time. Therefore, the spectra shown for these models have been shifted by 1~s. } 
\label{LyC_spectra}
\end{figure*}

%The $1F11$,$\delta$=3, $E_c$=20keV and $3F10$, $E_c$=20keV, $\delta$=6--7 models show a secondary increase in LyC emission after the peak beam heating. This increase is caused by the merging of the optically thick LyC layer with an optically thin component that forms due to chromospheric condensation. The optically thin LyC layers are discussed in more depth in Section~\ref{CF}.
%discuss primiary decrease d5 1f11 model

\subsection{Synthetic LyC spectra} %\footnote{There is an issue with the \ion{Ca}{2} data points for the majority of models shown. This issue affects LyC for the $1F10$, $\delta$=5, $E_c$=15keV model and the $3F10$, $\delta$=4, $E_c$=20keV model. These models have been omitted from the rest of this analysis.}}
\label{spectra}

Figure~\ref{LyC_spectra} shows continuum spectra at the time of the peak LyC emission in each simulation, between 9.7~s and 13.6~s. The \ion{He}{1} edge at 503.98\AA\ is also visible. The top row shows spectra generated for a fixed low-energy cutoff of 20~keV, while increasing the beam flux from $3F9$ to $1F11$ (left to right), and varying the spectral indices from $\delta=3$--$7$ (colored lines). The pre-flare spectra are shown in black. There were some snapshots ($<1$~s) in a few simulations in which the EUV continua tails had negative intensities, likely due to numerical noise. The simulations where the LyC was affected are: the $3F10$, $\delta$=5, $E_{c} = 20$~keV model and the $1F10$, $\delta$=5, $E_{c} = 20$~keV model.  Therefore, the spectra shown for these models have been shifted by 1~s. The spectra for the models where only the \ion{Ca}{2} continuum was affected were not shifted. As the electron beam intensifies, so too does the response of LyC, peaking roughly co-temporally with the peak of the heating rate. The $1F11$, $\delta=4$, 20keV model is an exception, peaking at a time of 13.6s as will be discussed in Section~\ref{lightcurves}. For the $3F9$ models, the head of the continuum is around four orders of magnitude greater than the pre-flare value. This increases to around 5.5 orders of magnitude for the $1F11$ models.  The magnitude of the LyC enhancement during solar flares is strongly dependent on the flux of non-thermal electrons. The dependence on the spectral index of the beam is weaker; softer beams result in slightly higher intensities. This is because softer beams (larger $\delta$) deposit their energy over a narrower region in the upper chromosphere since they thermalize more easily, resulting in more localized heating and electron density enhancements. 

%collisions with the plasma where the optically thick LyC layer forms.% \gsk{GSK: is it really just the collisions with the beam... the atmospheric response will also be very different and we can't ignore that}

The bottom row of Figure~\ref{LyC_spectra} shows synthetic LyC spectra for a fixed spectral index of $\delta = 5$ and $E_c = [15, 20, 25]$~keV (colored lines). The LyC spectra are weakly dependent on the low-energy cutoff, similar to the behaviour seen with varying the spectral index; Beams with larger $E_c$ values are composed of a greater proportion of high-energy electrons that penetrate deeper into the atmosphere, resulting in the beam heating region shifting downwards, affecting the LyC response. The increase in intensity from varying the spectral index and low-energy cutoff is much weaker compared with that caused by the increasing flux of the non-thermal electrons.

Like \citet{Ding1997}, we found a strong dependence of the LyC spectral intensity upon the flux of the non-thermal electron beam, as seen in Figures~\ref{LyC_spectra}. However, RADYN predicts larger impulsive phase LyC intensities than the modelling of \citet{Ding1997}. \citet{Ding1997} also predicted a stronger dependence on the spectral index than RADYN. These differences may arise for a number of reasons. Chiefly, \cite{Ding1997} used a static code with semi-empirical atmospheres and manually varied the temperature and transition region location. The electron beam was only used to add non-thermal collisions, and did not drive the chromospheric response. In our dynamic simulations, while these non-thermal collisions certainly play an important role, their impact on driving the overall dynamics is taken into account, so that there is a feedback between the atmosphere and the beam propagation/thermalisation.

Recent modelling of LyC emission during solar flares was also performed by \citet{Druett2019} using the radiative hydrodynamic code, HYDRO2GEN. They present LyC spectra for their 1F10, 1F11, and 1F12 solar flare models with $E_c = 10$~keV. The magnitude of our impulsive phase LyC intensities are in agreement with their spectra. However, the F-CHROMA grid does not include any 1F12 simulations.

%need to talk to graham about this now the E_c values have been included showing negative intensities in the LyC as this may be the same issue

\subsection{Synthetic LyC lightcurves}\label{lightcurves}

\begin{figure*}[ht!]
\includegraphics[width=1\textwidth]{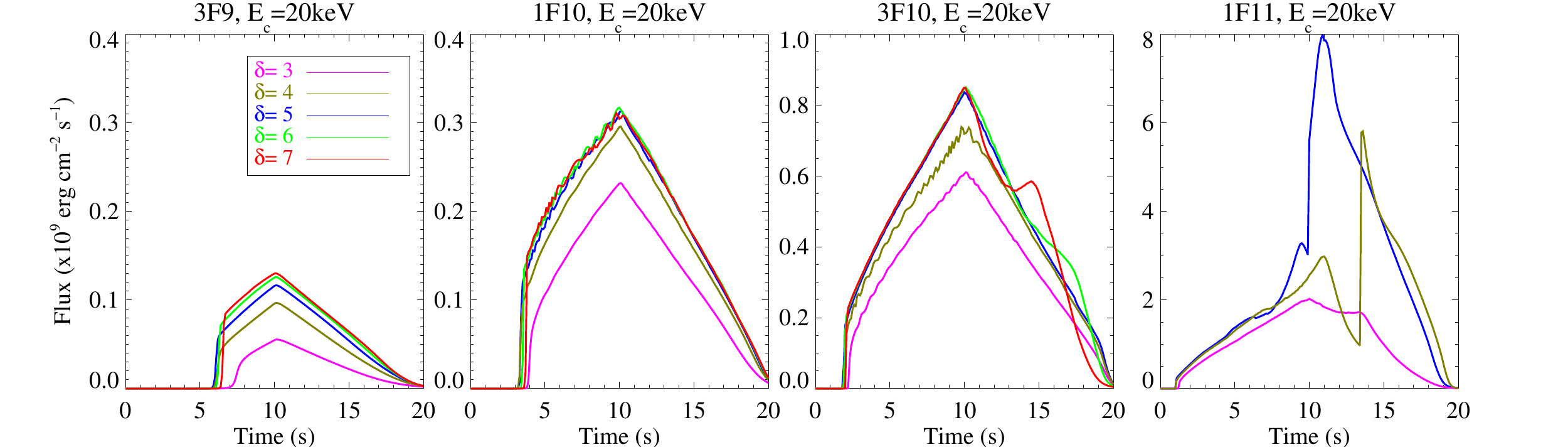}
\caption{Flare excess lightcurves of spectrally integrated LyC emission from RADYN (505.0\AA\ to 911.7\AA) for the $3F9$, $1F10$, $3F10$, and $1F11$ models with $\delta$=$3$--$7$ and $E_c$=20keV.% The background LyC flux was assumed to be the flux at t=0s and was subtracted at every time step. 
%The insert shows the light curve dimming that occurs between t=$1.7$--$1.9$s for the 3F11, $\delta$=7, and $E_c$=20keV model.
}
\label{LyC_lightcurves}
\end{figure*}

The flare excess integrated LyC lightcurves are shown in Figure~\ref{LyC_lightcurves} for the $3F9$, $1F10$, $3F10$, and $1F11$ models, with $E_c = 20$~ keV, and spectral indices ranging from $\delta = 3$--$7$ (colored lines). Note the saw-toothing seen for some of the models is due to numerical noise in the spectra between time steps. The synthetic lightcurves again highlight that the magnitude of the LyC enhancement is primarily driven by the magnitude of the non-thermal electron flux, whereas variations in the spectral index are less significant. Softer beams result in slightly higher LyC fluxes. This is because softer beams (larger $\delta$) deposit their energy over a narrower region in the upper chromosphere, resulting in more localized heating close to the LyC formation region. There is a period of time near the beam onset where the LyC emission is barely enhanced (recall that the flux of the non-thermal electron distribution ramps up in a triangular profile). The duration of this period is dependent upon the flux of the non-thermal electron beam, with larger fluxes having a shorter delay in the enhancement. All models consistently showed that notable LyC enhancement occurred after approximately $3\times10^{9}$--$4\times10^{9}$~erg~cm$^{-2}$ of energy was deposited by the beam. Flares that inject larger non-thermal electron fluxes more rapidly produce conditions conducive to LyC ionizations.

Despite this variable onset time, each LyC flare reaches a maximum around $t\sim10$~s and falls thereafter when the energy flux of the electrons decreases. The LyC response largely tracks the beam heating profile, in agreement with \cite{Druett2019}. However, the $1F11$, $\delta = 4$, $E_c = 20$~keV model shows a secondary peak in the LyC lightcurve at $t = 13.6$~s. This second maximum occurs as a chromospheric `bubble' (that is, a region of dense, low-temperature material, appearing at high altitude) dissipates at that time, resulting in a decrease in the optical depth \citep{Reid2020}. At the same time, a condensation front producing optically thin LyC emission merges with the heated chromosphere producing the optically thick LyC layer, resulting in a sudden, secondary increase in the LyC intensity. The optically thin LyC layers are discussed in more detail in Section~\ref{CF}. 

The $1F11$, $\delta = 3$, $E_c = 20$~keV and $3F10$, $E_c = 20$~keV, $\delta = 6,7$ models also show a secondary increase in LyC emission after the peak beam heating. This is caused by the merging of optically thin condensation fronts with the optically thick LyC layer. However, for these models, the chromospheric bubbles do not dissipate before this merging. For the $1F11$, $\delta = 5$, $E_c = 20$~keV model the merging of the two layers occurs before the peak beam heating, resulting in the peak seen between $9$--$10$~s in Figure~\ref{LyC_lightcurves} (the formation layers for this model are discussed in further detail in Section~\ref{CF}).

All this is to say that features in the lightcurves, while largely following the profiles of energy injection, are influenced by the complex dynamics of the flaring chromosphere. 

\subsection{LyC Formation Properties}\label{CF}
\subsubsection{Formation Heights and Optical Depth}

\begin{figure*}[ht!]
\includegraphics[width=1\textwidth]{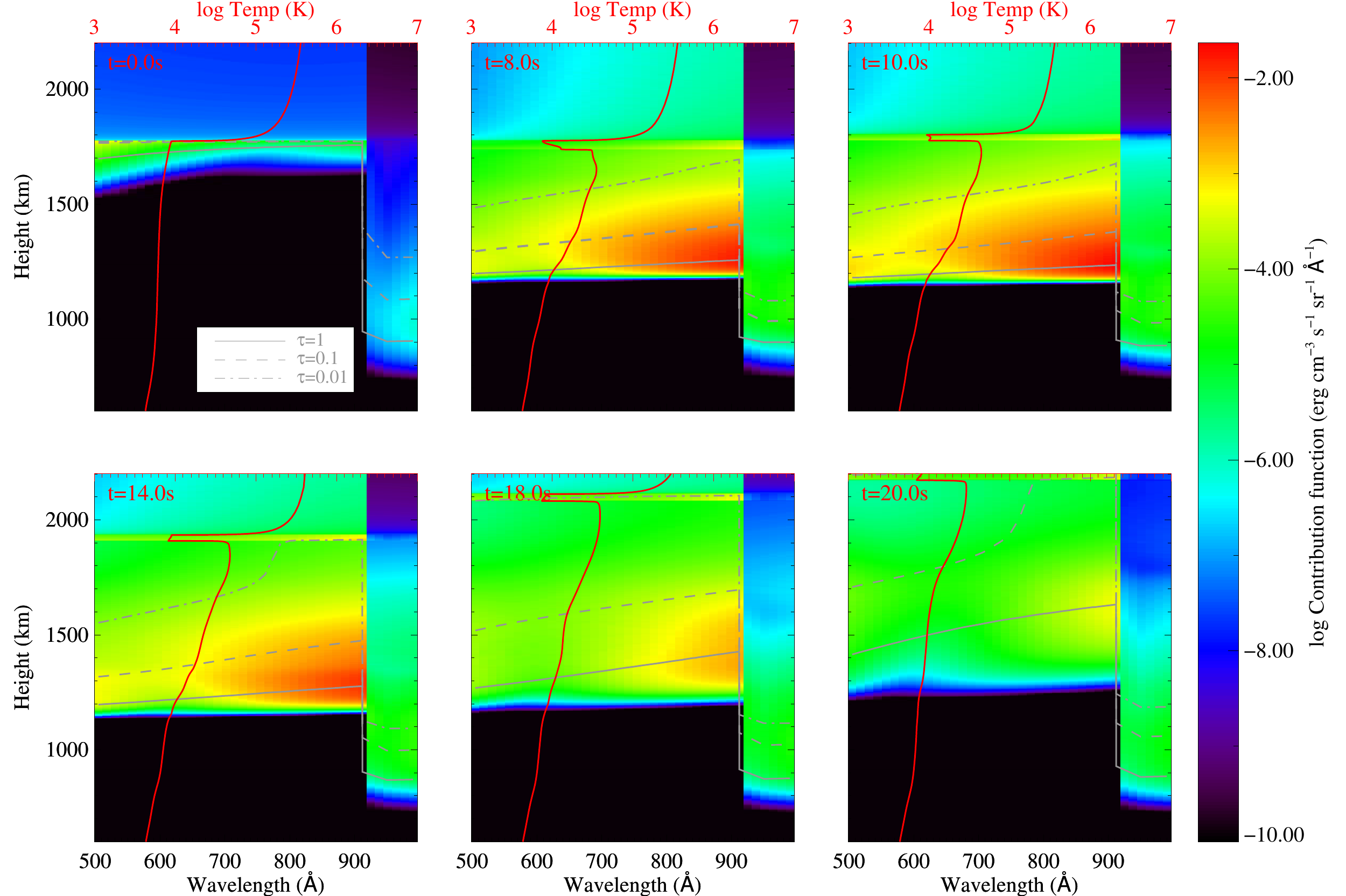}
\caption{The log of the contribution function for the $3F9$, $\delta = 5$, $E_c = 20$~keV model at times $t = [0, 8, 10, 14, 18, 20]$~s. The heights at which $\tau$=1 (grey solid line), $\tau$=0.1 (grey dashed line), and $\tau$=0.01 (grey dot-dashed line) are shown. The temperature profile is shown in red at each time step.}
\label{contrib_3e10_d5}
\end{figure*}

\begin{figure*}[ht!]
\includegraphics[width=1\textwidth]{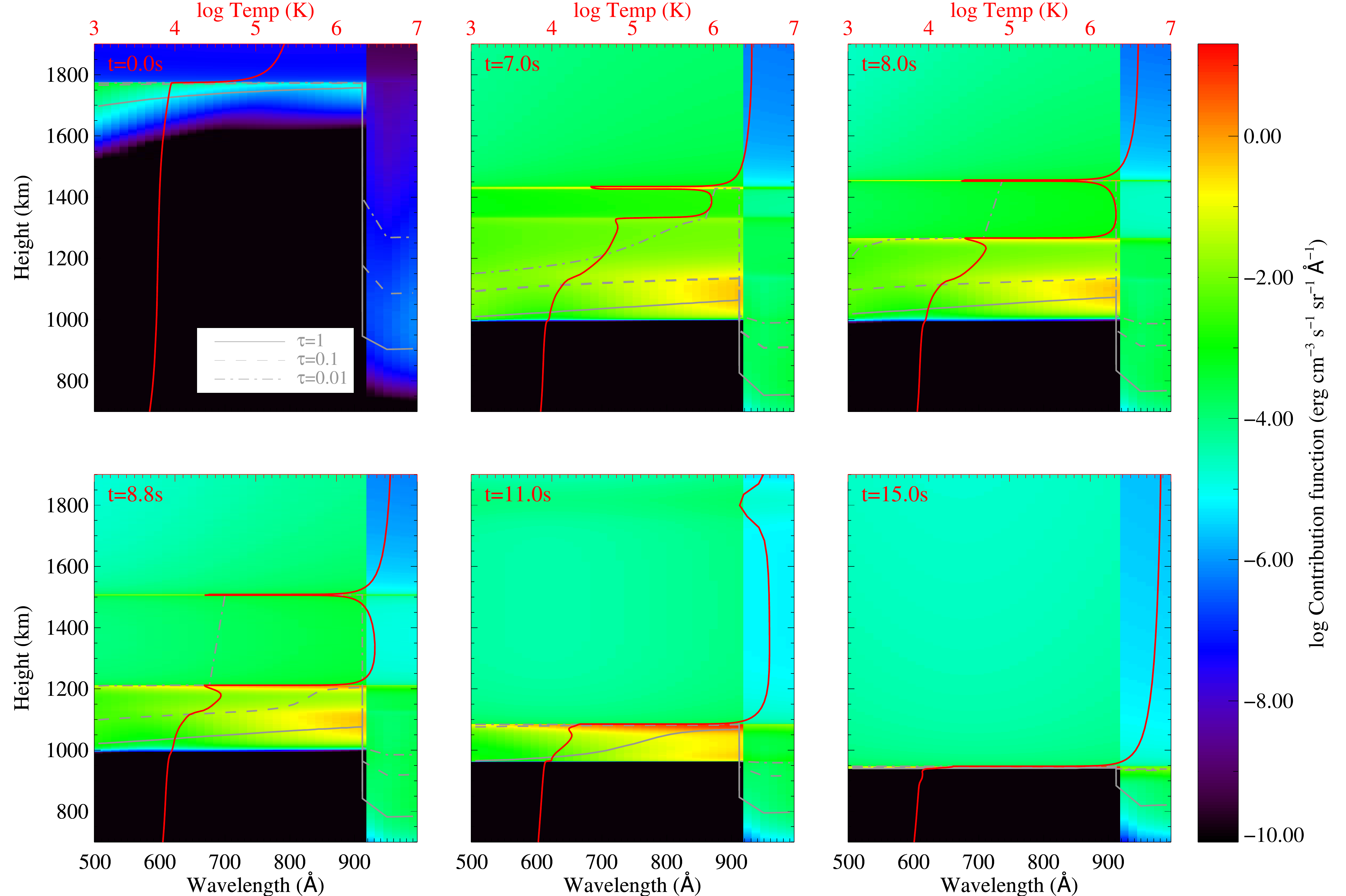}
\caption{Same as Figure~\ref{contrib_3e10_d5}, but for the $1F11$, $\delta = 5$, $E_c = 20$~keV model. Note the color bar scale has changed. }
\label{contrib_1e11_d5}
\end{figure*}

The contribution function to the emergent intensity effectively shows where in the solar atmosphere the emission originates (for brevity we refer to this simply as the `contribution function' going forward). The formal solution of the emergent intensity can be expressed as:

% \begin{equation}
% I_\lambda=\frac{1}{\mu} \int_z S_\lambda \frac{\chi_\lambda}{\tau_\lambda} \tau_\lambda \exp\biggl({\frac{-\tau_\lambda}{\mu}\biggr)} \mathrm{d}z = \int_z C_{I}~\mathrm{d}z,
% \label{EQN:contrib}
% \end{equation}

\begin{equation}
I_\lambda = \frac{1}{\mu} \int_{z} C_{I}(z) \mathrm{d}z = \int_{z} j_{\lambda}\exp{(\tau_{\lambda}/\mu}) \mathrm{d}z
\label{EQN:contrib}
\end{equation}

\noindent where the integrand is the contribution function, $C_{I}(z)$ \citep{1986A&A...163..135M,1994chdy.conf...47C}. Following \cite{2017ApJ...836...12K} we express $C_{I}(z)$ in terms of emissivity, $j_{\lambda}$, attenuated by some optical depth $\tau_{\lambda}$. The emissivity and opacities are the sum of various sources, including hydrogen bound-free, hydrogen free-free, H$^{-}$, scattering by metals, Rayleigh scattering, and Thomson scattering. Of these, the bound-free transitions are treated fully in NLTE, whereas the others are treated in LTE, but using the NLTE, non-equilibrium electron or proton densities as appropriate (See \cite{2017ApJ...836...12K} for more details).

%In this expression, $\chi_\lambda$ is the monochromatic opacity (also referred to as the linear extinction coefficient), $\tau_{\nu}$ is the optical depth, $S_\nu$ is the source function given by the ratio of the emissivity to the extinction coefficient, and $\mu$ is the cosine of the viewing angle.  %\gsk{GSK: I will rewrite this in terms of the emissivity, which might make more sense since we mainly discuss emissivity later, not the $\chi/\tau$ type terms... Eq 3 as it currently is written was cast in that way for understanding the spectral line formation via Mats' 4-panel style figs... for the continuum we aren't looking at velocity gradients and the impact on line shape etc., so there is no point in having $\chi/\tau \tau$} 

Generally, if the emission originates close to the $\tau_{\nu} = 1$ surface then the emission forms under optically thick conditions, and if it originates in a region where $\tau_{\nu}<<1$ then it can be considered to be optically thin. It is possible that some flares may exhibit both optically thick and thin components with significant emission coming from multiple layers, e.g. from an optically thin layer overlying the chromosphere. Knowledge of the plasma properties in the LyC forming region(s), in response to different heating conditions, will broaden our understanding of energy transport processes during flares, and where this energy is deposited. %\gsk{GSK: can we make a stronger statement here... something like leading to potential diagnostics of the flaring plasma (b1, temperature etc.,) or being able to determine that there are multiple layers based on LyC spectra (which is important as we think there are multiple components to spectral lines etc., etc.,}

%For the $3F9$, $\delta$=5, $E_c$=20keV and the $1F11$, $\delta$=5, $E_c$=20keV models the total contribution function and the various sources of emissivity and opacity that produced the emergent intensity were investigated. These two cases where selected to investigate the LyC formation for a less energetic and more energetic solar flare of the 25 models used. The hardness of the beam was varied to investigate the dependence of the LyC formation height for different energy deposition sites. 
To illustrate the general formation properties we discuss detailed examples from two simulations: the $3F9$, $\delta = 5$, $E_c = 20$~keV and the $1F11$, $\delta = 5$, $E_c = 20$~keV models. These showcase a weaker and stronger flare, respectively. Figure~\ref{contrib_3e10_d5} shows the temporal evolution of the log of the contribution function for the $3F9$, $\delta = 5$, $E_c = 20$~keV model during the heating phase. The temperature stratification (solid red line) and the heights at which $\tau_{\lambda}=[0.01, 0.1, 1]$ are also indicated (grey dot-dashed, dashed, and solid lines, respectively). Initially, at $t = 0$~s, there is a peak in the contribution function at the base of the transition region (z $\sim1750$~km), forming between $0.1 \lesssim  \tau_{\lambda} \lesssim 1$, corresponding to an optically thick LyC formation, in agreement with the literature \citep{Machado1978,Machado2018,Druett2019}. By $t = 8$~s, the formation region has widened, and shifted deeper into the chromosphere ($z\approx1200-1400$~km) as the beam heating strengthens. The height difference between $\tau_{\lambda} = 1$ for short and long wavelengths has also narrowed. At this time, opacity effects are still significant, with the bulk of the emission forming between $0.1 \lesssim  \tau_{\lambda} \lesssim 1$. A secondary peak in the contribution function is present higher in altitude ($z= 1700$~km), co-spatial with a chromospheric `bubble', seen as a narrow dip in the temperature profile forming at the base of the transition region at $t=8$, $10$, $14$, $18$, and $20$~s in Figure~\ref{contrib_3e10_d5}. Emission from this very narrow layer forms under optically thin conditions $\tau_{\lambda}< 0.01$. Similar regions of cooler, yet dense plasma, are present in most of our simulations \citep[see also][]{Allred_2005,Reid2020}. \cite{Druett2019} also presented evidence for an overlying optically thin layer in their 1F11, $E_c$=10keV model. As time progresses, this chromospheric `bubble' propagates upwards towards the corona along a chromospheric ablation (also referred to as `evaporation') front. The bubble collapses as it propagates, narrowing until all of the cool material has been heated to coronal temperatures (or higher). In the declining phase of the heating ($t>10$~s) the $\tau_{\lambda} = 1$ layer begins to gradually return to the pre-flare height as the atmosphere cools and the number of recombinations in low opacity regions increases whilst ionization decreases, in agreement with \citep{Druett2019}. The $\tau_{\lambda} = 0.1$ layer begins shifting to the location of the bubble as the density there increases, raising the LyC opacity at those altitudes.

%\cite{Druett2019} also presented evidence for an optically thin layer in their 1F11, $E_c$=10keV model. At later times, the optically thin layers dissipated in our simulations and the optically thick layer began to relax back to its pre-flare height, again in agreement with their findings.

Figure~\ref{contrib_1e11_d5} shows the temporal evolution of the log of the LyC contribution function for the $1F11$, $\delta = 5$, $E_c = 20$~keV model during the heating phase. In this more energetic simulation, the LyC forms over much more narrow regions ($z\approx 1000$--$1100$~km) due to the more dramatic response of the chromosphere that becomes very compressed. By $t = 8$~s, two optically thin layers have formed due to bubbles flanking a rapidly expanding high-temperature region (one upflowing bubble due to explosive evaporation at $z\approx1450$~km, and one downflowing bubble at $z\approx1250$~km, commonly referred to as a chromospheric condensation). When the very dense downflowing bubble effectively merges with the bulk of the chromosphere the density at temperatures favorable to produce LyC can rapidly increase. Since opacity is a strong function of wavelength there can be times at which there is a marked optical depth stratification, resulting in variations to the spectral shape of the LyC (e.g. flattening at certain wavelengths). Examples are shown at $t = [8.8, 11]$~s, where the $\tau_{\lambda} = 1$ layer forms at a higher altitude for the head of the continuum compared to the tail. In this stronger flare, the chromosphere becomes so compressed that the LyC ultimately forms from a vanishingly narrow downflowing layer in the latter stages of the flare, shown at $t=15$~s of Figure~\ref{contrib_1e11_d5} at $z\approx1000$~km. %Lastly, when the energy injection has ceased, the atmosphere is still sufficiently hot and dense that the optically thick LyC layer can not begin relaxing back to its preflare state until much later in the simulation. %The $1F11$, $\delta$=5, $E_c$=20keV model also forms an optically thin bubble before t=8s, however, it dissipates by t=7s so is not shown in Figure~\ref{contrib_1e11_d5}. 

% \begin{figure}[ht!]
% \includegraphics[width=0.5\textwidth]{LyC_OpacEmissEx_1F11d5Ec20_varytimes_wref1.pdf}
% \caption{The components of $\lambda = 900$~\AA\ opacity (left column) and emissivity (right column) for four snapshots in the $1F11$, $\delta = 5$, $E_c = 20$~keV model. The colored lines correspond to various contributions to the totals shown as black, thick lines. The vertical lines correspond to the heights at which $\tau$=1 (dashed), $\tau$=0.1 (dotted), and $\tau$=0.01 (dot-dashed).}
% \label{fig:emiss_wref1}
% \end{figure}

These formation properties are generally true of the other simulations. Most simulations contain some combination of optically thick and overlying optically thin contributions to the emergent spectra (though the bulk of the emission does tend to originate between $0.1 \lesssim  \tau_{\lambda} \lesssim 1$). Stronger flares produce more dramatic atmospheric responses, leading to the bubbles and shocks, and associated phenomenon. These dynamics also emerge and develop faster with increasing flare strength, and for softer non-thermal electron distributions. This is because these electrons heat the uppermost chromosphere very efficiently and do not penetrate as deeply, driving flows more easily. Weaker flares and harder spectra do not tend to produce as many shocks, do not significantly compress the chromosphere, and exhibit mainly upflowing features.

%\begin{figure*}[ht!]
%\includegraphics[width=0.75\textwidth]{LyC_OpacEmissEx_1F1%1d5Ec20_varytimes_wref1.pdf}
%\caption{The components of $\lambda = 900$~\AA\ opacity (left column) and emissivity (right column) for four snapshots in the $1F11$, $\delta = 5$, $E_c = 20$~keV model. The colored lines correspond to various contributions to the totals shown as black, thick lines. The vertical lines correspond to the heights at which $\tau$=1 (dashed), $\tau$=0.1 (dotted), and $\tau$=0.01 (dot-dashed).}
%\label{fig:emiss_wref1}
%\end{figure*}

%\begin{figure*}[ht!]
%\includegraphics[width=0.75\textwidth]{LyC_OpacEmissEx_1F11d5Ec20_varytimes_wref2.pdf}
%\caption{Same as Figure~\ref{fig:emiss_wref2} but for $\lambda = 600$~\AA.}
%\label{fig:emiss_wref2}
%\end{figure*}

\subsubsection{Emissivity and Opacity} \label{Emiss_Opac_section}

% \begin{figure*}[ht!]
% \includegraphics[width=0.75\textwidth]{image001.png}
% \caption{The components of $\lambda = [700, 900]$~\AA\ opacity (top row) and emissivity (bottom row) for two snapshots in the $1F11$, $\delta = 5$, $E_c = 20$~keV model. The light grey dashed line shows the total opacity or emissivity at t=0s in each panel. The colored lines correspond to various contributions to the totals shown as black, thick lines. The vertical lines correspond to the heights at which $\tau$=1 (dashed), $\tau$=0.1 (dotted), and $\tau$=0.01 (dot-dashed).}
% \label{fig:emiss_wref}
% \end{figure*}

\begin{figure*}[ht!]
\includegraphics[width=\textwidth]{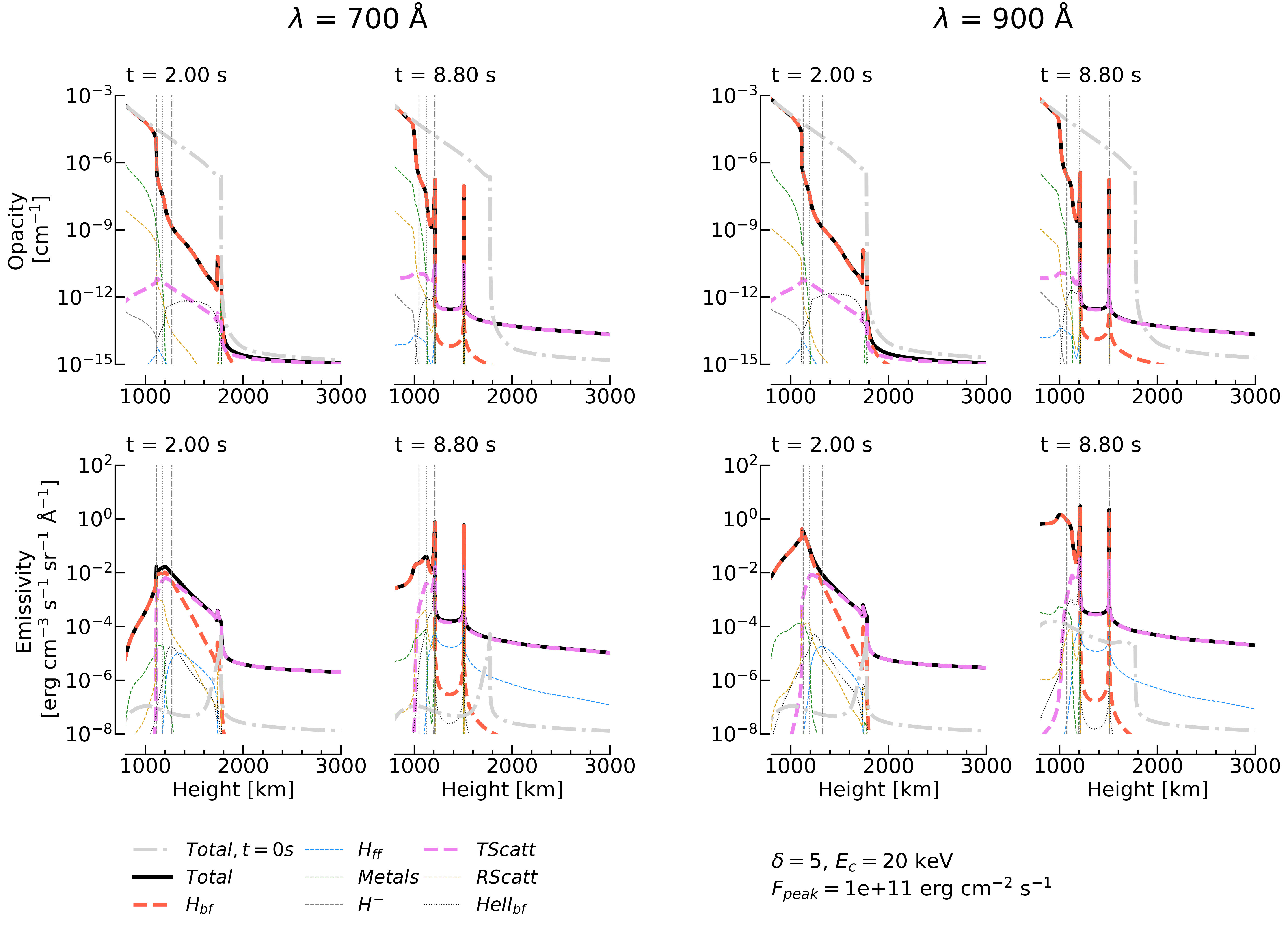}
\caption{The components of $\lambda = [700, 900]$~\AA\ opacity (top row) and emissivity (bottom row) for two snapshots in the $1F11$, $\delta = 5$, $E_c = 20$~keV model. The left-hand panels show $700$~\AA\ and the right-hand panels show $900$~\AA. The light grey dot-dashed line in each panel shows the total opacity or emissivity at $t = 0$~s. The colored lines correspond to various contributions to the total emissivity or opacity (the sums of these contributions are shown as thick black lines). The vertical lines correspond to the heights at which $\tau = 1$ (dashed), $\tau = 0.1$ (dotted), and $\tau = 0.01$ (dot-dashed).}
\label{fig:emiss_wref}
\end{figure*}

To understand some of the features described we can briefly look at the components of opacity and emissivity as functions of space, time, and wavelength. We illustrate the components of emissivity and opacity at two times from one flare for a long ($\lambda = 900$~\AA) and short ($700$~\AA) wavelengths in Figures~\ref{fig:emiss_wref}. The dominant source of opacity at all wavelengths, throughout the chromosphere and through the lower transition region was hydrogen free-bound opacity, $\chi_{\lambda,H_{bf}}$ (red dashed line). Towards the upper transition region and corona, where hydrogen is ionized, Thomson scattering begins to dominate but the overall effect on the emerging radiation is small, compared to the free-bound opacity (pink dashed line). %However outside of the dense cool bubbles, the opacity in the upper atmosphere is significantly smaller than the chromosphere. 

Once the flare starts in earnest, we saw that the formation height of the LyC dropped in altitude. This is because the upper chromosphere was strongly heated, ionizing hydrogen and decreasing $\chi_{\lambda,H_{bf}}$, meaning that the $\tau_{\lambda}  = 1$ forms much deeper in the atmosphere, shown in the top panels of Figure~\ref{fig:emiss_wref}. Later, the merging of the dense bubbles with the bulk of the chromosphere means that $\chi_{\lambda,H_{bf}}$ increases, but since this is a function of wavelength, this happens first for longer wavelengths during this process.

The emissivity of the LyC is also dominated by hydrogen recombinations, $j_{\lambda,H_{bf}}$ at longer wavelengths (roughly speaking $\lambda > 700$~\AA), but at shorter wavelengths Thomson scattering can be important. Thomson scattering can also compete with $j_{\lambda,H_{bf}}$ at longer wavelengths at certain times, but once the electron density is high (e.g. in the narrow chromospheric bubbles) then $j_{\lambda,H_{bf}}$ once again dominates. The more extended region of flaring LyC formation compared to the quiet Sun can be understood from the stratification of emissivity and opacity also. During a flare, a greater extent of the chromosphere is at an elevated temperature, and hence electron density, so the free-bound emissivity throughout the greater geometric height range is raised. Integrating over height yields a higher emergent intensity. 

In Figures~\ref{fig:emiss_wref}, note the initial decrease in opacity (top row t=2s panels for 700\AA\ and 900\AA), pushing LyC formation deeper, and then the influence of the bubbles. The emissivity is largest at the optically thick layer (between $\tau=0.1$--$1$ heights) as the density of emitting particles is greatest here. The optically thin bubbles also have an increased emissivity compared to the ambient plasma. At shorter wavelengths the emissivity of the optically thicker layer is comparable to the emissivity of the optically thin bubbles, meaning the optically thin layers will reinforce the lower wavelength LyC intensities to a greater extent than the continuum head, in agreement with the literature \citep{Machado1978, Machado2018,Druett2019}.

\section{LyC color Temperature and Departure from LTE}\label{b1_analysis}
\subsection{Fitting the LyC spectra}\label{LyC_b1_spectra}

\begin{figure*}[t]
\centering
\includegraphics[width=\textwidth]{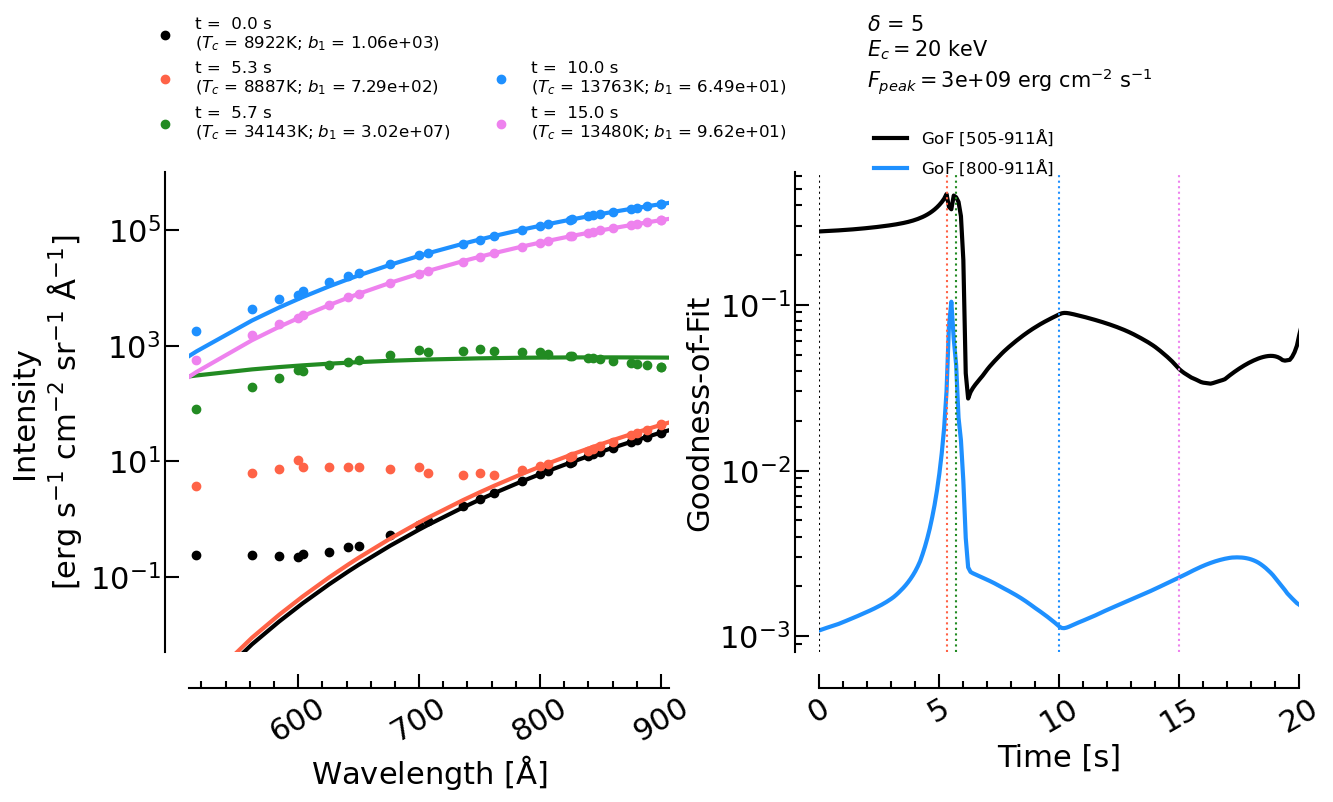}
\caption{Left-hand panel shows synthetic RADYN spectra from 400--911\AA\ for the $3F9$, $\delta$=5, $E_c$=20keV model, at times t=0s, 5.3s, 5.7s, 10s, and 20s (colored dots). Fits to LyC using Equation~\ref{EQN:E-B} are shown by solid colored lines. The right-hand panel shows the log of the goodness-of-fit ($\epsilon$) as a function of time for the fits shown in the left panel (solid black curve), while the blue curve illustrates the $\epsilon$ values when the data was fit between $800.0$\AA\ and $911$\AA. The dashed colored lines denote the times of the spectra in the left panel. % The \ion{He}{1} edge at 503.98\AA\ and LyC edge at 911.12\AA\ are denoted by vertical green dashed lines), while the tail of the \ion{Ca}{2} continuum ($<$1044.00\AA) is also visible.}
}
%\gsk{GSK: graham to think more about this.. prob C and Si, and other metals, really...}
\label{Example_EB_fit}
\end{figure*}

%Given that LyC is sensitive to the energy balance in chromospheric structures, and since thermalization occurs very rapidly at the higher densities located here, the LyC spectrum can reflect the local plasma temperature. LyC could thus provide a crucial diagnostic for understanding energy deposition in the chromosphere during solar flares, and plasma conditions therein.

 Following the method used by \cite{Machado1978} and \cite{Machado2018}, the synthetic LyC spectra were fit using the Eddington-Barbier (EB) relation to determine values for $b_1$ and $T_c$. This approximation equates the emergent intensity to the source function at optical depth unity, and is given by:
%which assumes that LyC can be approximated by a blackbody function.
\begin{equation}
I_{\lambda}(\mu)\approx S_{\lambda}(\tau_{\lambda}=\mu)=\frac{B_\lambda(T_c)}{b_1},
\label{EQN:E-B}
\end{equation}

\noindent
where $B_{\lambda}(T_c)$ is the Planck function,

\begin{equation}
B_{\lambda}(T_c)=\frac{2hc^2}{\lambda^5} \frac{1}{\exp\left(\frac{hc}{\lambda k_B T_c}\right)-1}.
\label{EQN:Planck}
\end{equation}

\noindent
$I_{\lambda}$ is the continuum intensity and $S_{\lambda}$ is the source function. The $\lambda$ subscripts indicate that these variables are functions of wavelength. $T_c$ is the color temperature, $b_1$ is the non-local thermodynamic equilibrium departure coefficient of the first level of hydrogen, and all other constants have their usual meanings.

An illustration of this spectral fitting is shown in the left-hand panel of  Figure~\ref{Example_EB_fit}, for various times during the $3F9$, $\delta = 5$, $E_c = 20$~keV model. The right-hand panel shows the temporal evolution of the log of the goodness-of-fit ($\epsilon$) with vertical dashed lines denoting the time of the spectra in the left panel. The goodness-of-fit values are given by:

\begin{equation}
\epsilon=\frac{1}{N} \sum_{i=1}^{N} \frac{|y_i-O_i|}{y_i},
\label{diff_sq}
\end{equation}

\noindent where $i$ is the element index, $N$ is the number of data points, $y_i$ is the data value and $O_i$ is the fitted value. As the goodness-of-fit is weighted by the data for each wavelength and time, smaller values correspond to a smaller relative difference between the data and the fit, meaning the fits are better.

%Note that the t=0s continuum tail falls below 0 on the logarithmic scale. Therefore, the large differences between the LyC tail and the $505.0$--$911.3$\AA\ fit correspond to fractional differences in real space. Therefore, over-fitting of the LyC occurs for both lines in the left-hand panel of Figure~\ref{Example_EB_fit}, resulting in the SDS values of less than one \gsk{GSK: we need to think about these statements, as its not immediately clear or intuituve how to interpret the sum of differences, and if your discussion of over-fitting is valid... I think normalising by $\Sigma y_i^2$, so that we have some fractional difference between the data and the fit might make more sense??}.

%At t=5.3s, the head of the continuum responds weakly to the non-thermal electron beam, whilst the tail significantly increases. This occurs as there is a narrow region of plasma between the beam heating region and transition region that is minimally heated, maintaining a large hydrogen ground state population compared to the ambient plasma. This pocket of plasma absorbs photons from the optically thick LyC layer formed below, resulting in the weak response of the LyC head near the beam onset. The same phenomenon causes the t=5.7s spectral profile. 

In the right-hand panel of Figure~\ref{Example_EB_fit}, at $t \sim 5$~s the beam has mostly heated a region in the atmosphere below the formation height of the LyC (i.e below $\tau_\lambda = 1$, where photons cannot readily escape). As discussed in Section~\ref{CF}, the height at which $\tau_\lambda = 1$ varies with wavelength, typically forming lower in the atmosphere with decreasing wavelength. The difference is a few tens of km at $t=5$~s. The flare does produce a small increase in emissivity above the height at which $\tau_\lambda = 1$ for shorter wavelengths, increasing the emergent intensity towards the tail of the LyC. At that height, however, $\tau_\lambda > 1$ for the head of the continuum and so there is no meaningful change in the emergent intensity for longer wavelengths. During these times, the LyC spectrum cannot be approximated by the EB assumption, resulting in the large spike in $\epsilon$ seen between $t = 5$--$6$~s in the right-hand panel of Figure~\ref{Example_EB_fit}, indicating a poor fit. Further discussion on this behaviour can be found in Section~\ref{SPIKES}.

In general, the tail of the continuum does not conform to the EB assumption as well as longer wavelengths, particularly in the decay phase following the flare peak. This is due to the presence of the overlying optically thin LyC layers discussed in Section~\ref{CF}. When Equation~\ref{EQN:E-B} is applied only between $\lambda = 800-911$\AA\ we obtain better fits (the blue line on the goodness-of-fit panel in Figure~\ref{Example_EB_fit}) compared to fitting the full range $\lambda = 505$--$911$\AA\ (black line). This suggests two distinct gradients of the LyC, in agreement with the literature \citep{Machado1978,Machado2018,Druett2019}. For times in the simulation where the fit was reasonable, we obtain $T_{c}$ values consistent with previous studies, increasing from $T_{c} \sim 9$~kK to $T_{c} \sim 13.5$~kK, with $b_{1}$ decreasing, though not reaching unity. Where we see a poor fit (e.g. $t=5.7$~s) we obtain an exceptionally large $b_{1}$ and $T_{c} > 34$~kK. 

\begin{figure*}[ht!]
\includegraphics[width=1\textwidth]{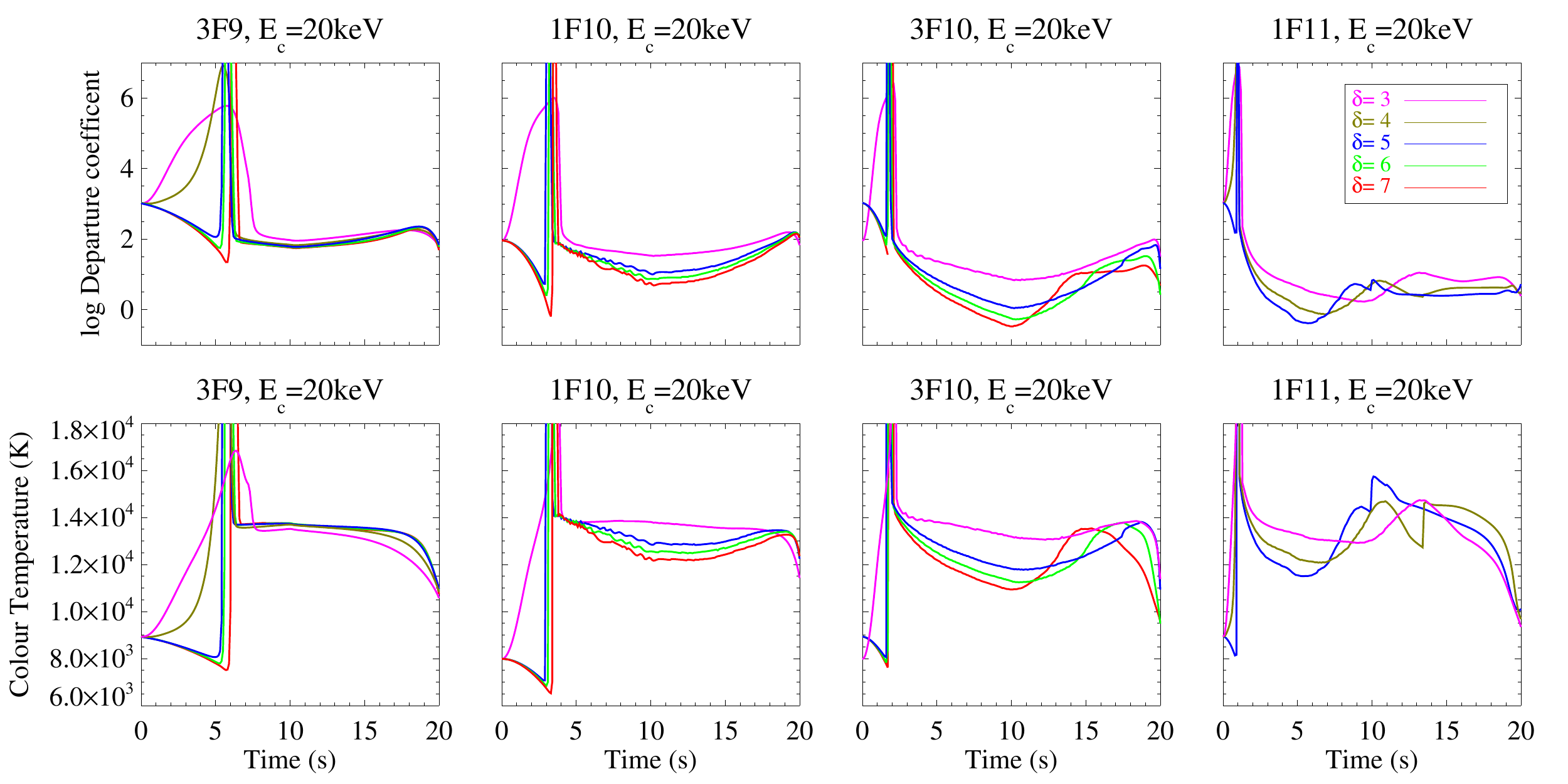}
\caption{The temporal evolution of $b_1$ and $T_c$ for the $3F9$, $1F10$, $3F10$, $1F11$ models with $E_c$=20keV and $\delta$=$3$--$7$.}
\label{b1_Tc_fits}
\end{figure*}

%\begin{figure}[ht!]
%\includegraphics[width=0.5\textwidth]{Temp_plot_thickness.eps}
%\caption{Temperature profiles for the $3F9$ models, with $E_c$=20keV, and $\delta$=3 and 5 at t=7s. The normalised beam heating is shown by the dashed lines for both models.}
%\label{Temp_profiles_spikes}
%\end{figure}

Values of $T_{c}$ and $b_{1}$ were obtained from fitting the LyC spectra by applying the EB assumption (Equations~\ref{EQN:E-B} and \ref{EQN:Planck}) at each time step from every flare in our study. Figure~\ref{b1_Tc_fits} shows the temporal evolution of $b_1$ (top row) and $T_c$ (bottom row) for the $3F9$, $1F10$, $3F10$, and $1F11$ models, with $\delta = [3$--$7]$ (colored lines), and $E_c = 20$~keV where we fit between $\lambda = 700$--$911$~\AA. This range was selected as Figure~\ref{Example_EB_fit} shows that the Eddington-Barbier approximation generally fits the data well at those wavelengths, but shorter wavelengths can deviate at certain times.
 
During the beam onset, the initial response of $b_1$ and $T_c$ is somewhat dependent on the spectral index. For beams with $\delta \geq 5$, $b_1$ initially decreases steadily during the beam onset followed by a rapid and sudden increase whereas, beams with $\delta \leq 4$ show an initial gradual increase in $b_1$, peaking around the time of the spikes in the other models. We limit the range of the figures to the scales shown, but the $b_1$ and $T_c$ values at the time of these spikes can be several orders of magnitude larger. 
 
The spikes in $b_1$ and $T_c$ are caused by the flattening in the continuum head, at which times the EB approximation is no longer a valid assumption; the $b_1$ and $T_c$ values at these times are unreliable. However, there are three key times when the behaviour of $b_1$ is consistent across all models. At $t = 0$~s, $b_1 \sim 10^{2}$--$10^3$, decreasing to a minimum around $t \sim 10$~s (flare peak), then increasing again during the decay phase. The magnitude of the $b_1$ minima depends upon the flux of the non-thermal electron beam: $b_1$ values for the $3F9$ models decrease from $b_{1} \sim 10^3$ to  $b_{1} \sim 10^2$, whereas the $3F10$ and $1F11$ models decrease closer $10^{-1} < b_{1} < 10^{1}$. The inference there being that more energetic simulations drive LyC formation closer to LTE due to the increased electron densities. Both thermal and non-thermal collisions will significantly increase in those simulations. \citet{Ding1997} presented a similar finding whereby beams with larger non-thermal electron fluxes caused $b_1$ to decrease to a greater extent. 

As discussed earlier, \citet{Machado2018} determined $b_1$ and $T_c$ values from pre-flare and flaring spectra for the six solar flares they analysed. The pre-flare fit results from our modelling are consistent with their quiescent $T_{c}$ and $b_{1}$ results. Further, the range of $T_{c}$ values we measured in our flare simulations are also generally consistent with the observations. During the observed flares, \cite{Machado2018} reported $b_{1} \approx 1$, whereas our model results show a wide range of values (between $0.1$--$10^2$) depending upon the flux of the non-thermal electron beam. One possible reason for this discrepancy may be due to the dependency of the observed $b_1$ values scaling with the assumed flaring area. Further details can be found in \citet{Machado2018}.

\begin{figure}
	\centering 
	\hbox{
		\subfloat{\includegraphics[width = 0.475\textwidth, clip = true, trim = 0cm 0cm 0cm 0cm]{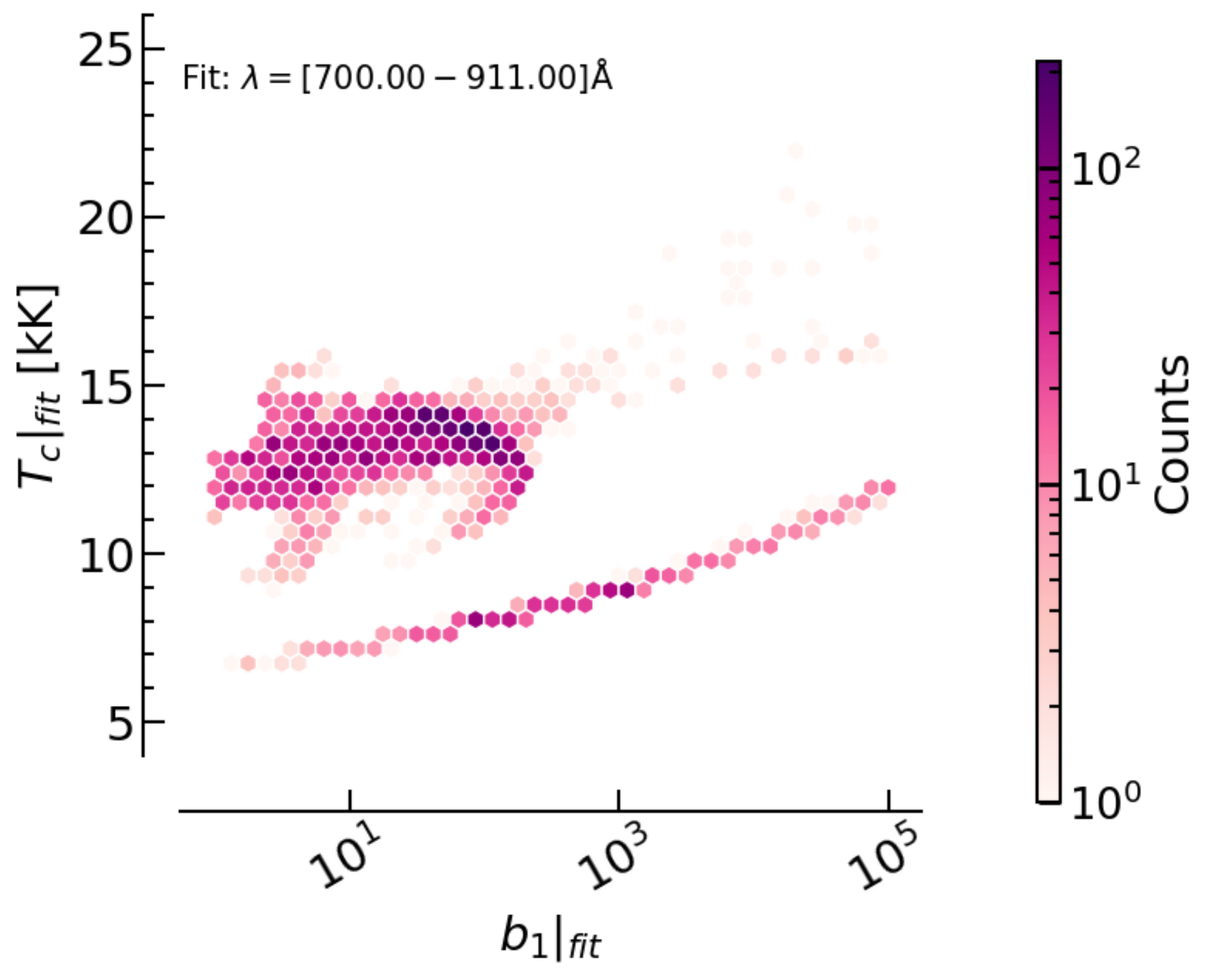}}
		}
    %\hbox{
	%	\subfloat{\includegraphics[width = 0.475\textwidth, clip = true, trim = 0cm 0cm 0cm 0cm]{FCRHOMA_HexHist_Tc_vs_b1_fit_700to911_TIME.pdf}}
	%	}
    %\hbox{
	%	\subfloat{\includegraphics[width = 0.475\textwidth, clip = true, trim = 0cm 0cm 0cm 0cm]{FCRHOMA_HexHist_Tc_vs_b1_fit_700to911_WEIGHTED.pdf}}
	%	}
	\caption{{Histogram of $T_{c}$ and $b_{1}$ obtained from fitting the EB approximation to our simulation grid, weighted by the number of counts.}}
	\label{fig:fit_histos}
\end{figure}

To visualise the spread of $T_{c}$ and $b_{1}$ values we produced a 2D histogram that collates the information from all simulations for all times with parameters: $\delta = [3,4,5]$, $E_{c} = [15, 20, 25]$~keV, $F_{peak} = [3F9, 1F10, 3F10, 1F11]$~erg~s$^{-1}$~cm$^{-2}$ (recall that the $F_{peak} = 1F11$ simulations did not include the $E_{c} = 15$~keV scenario). This is shown in Figure~\ref{fig:fit_histos}, where it is clear that $T_{c}$ increases from quiet Sun values, $T_{c} \sim 8$--$9$~kK, to values roughly in the range $T_{c} \sim 10-16$~kK, in agreement with the literature \citep{Machado1978,Ding1997,Lemarie_2004,Machado2018}. There are two distinct clusters within the histogram; the cooler cluster ($T_c\sim$ 6--11~kK, $b_1\sim10^1-10^5$) and the hotter cluster ($T_c\sim$ 10--20~kK, $b_1\sim10^1-10^3$). The cooler cluster corresponds to times before the brief breakdown in the EB approximation, whilst the hotter cluster corresponds to times after this.

%During the peak of the beam heating, as $b_1$ approaches unity, $T_c$ is approximately equal to the electron temperature of the plasma, $T_e$. Therefore the LyC forms at $T_e\approx12000$--$16000$K, as seen from Figure~\ref{b1_Tc_fits}. %\gsk{(GSK: Maybe this is present later but we can also look at the mean formation temperature directly and compare)}

\subsection{Comparing Spectral Fitting to Derived Plasma Properties}\label{model_fit_comparison}

We can assess how consistent the properties derived from fitting the spectra are with the actual plasma conditions in the models by comparing the $b_1$ values directly from RADYN, that we refer to as $b_{1,rad}$. The atomic level populations are functions of height and wavelength, and so to obtain $b_{1,rad}$ averaged over the LyC formation region we calculate the normalised cumulative distribution function (NCDF) of the contribution function, $C_{cdf}$ \citep[see also e.g.][]{2017ApJ...836...12K}. The heights corresponding to where the bulk of the emission originates are selected. The weighted average of $b_{1,rad}$ in that formation region was then obtained, weighted by the contribution function:

\begin{equation}
<b_{1,rad}> = \frac{\int_{z(C_{cdf}=low)}^{z(C_{cdf}=upp)}C_{I}(z)~b_{1,rad}~\mathrm{d}z}{\int_{z(C_{cdf}=low)}^{z(C_{cdf}=upp)}C_{I}(z)~\mathrm{d}z},
\label{eq:avvalue}
\end{equation}

\noindent where $z(C_{cdf}=low)$ refers to the height at which the $C_{cdf}$ reaches the lower bound, and $z(C_{cdf}=upp)$ the height of the upper bound. For example, the heights corresponding to $10\%$ \& $90\%$ of the $C_{cdf}$, respectively. 

\begin{figure}[ht!]
\includegraphics[width=0.5\textwidth]{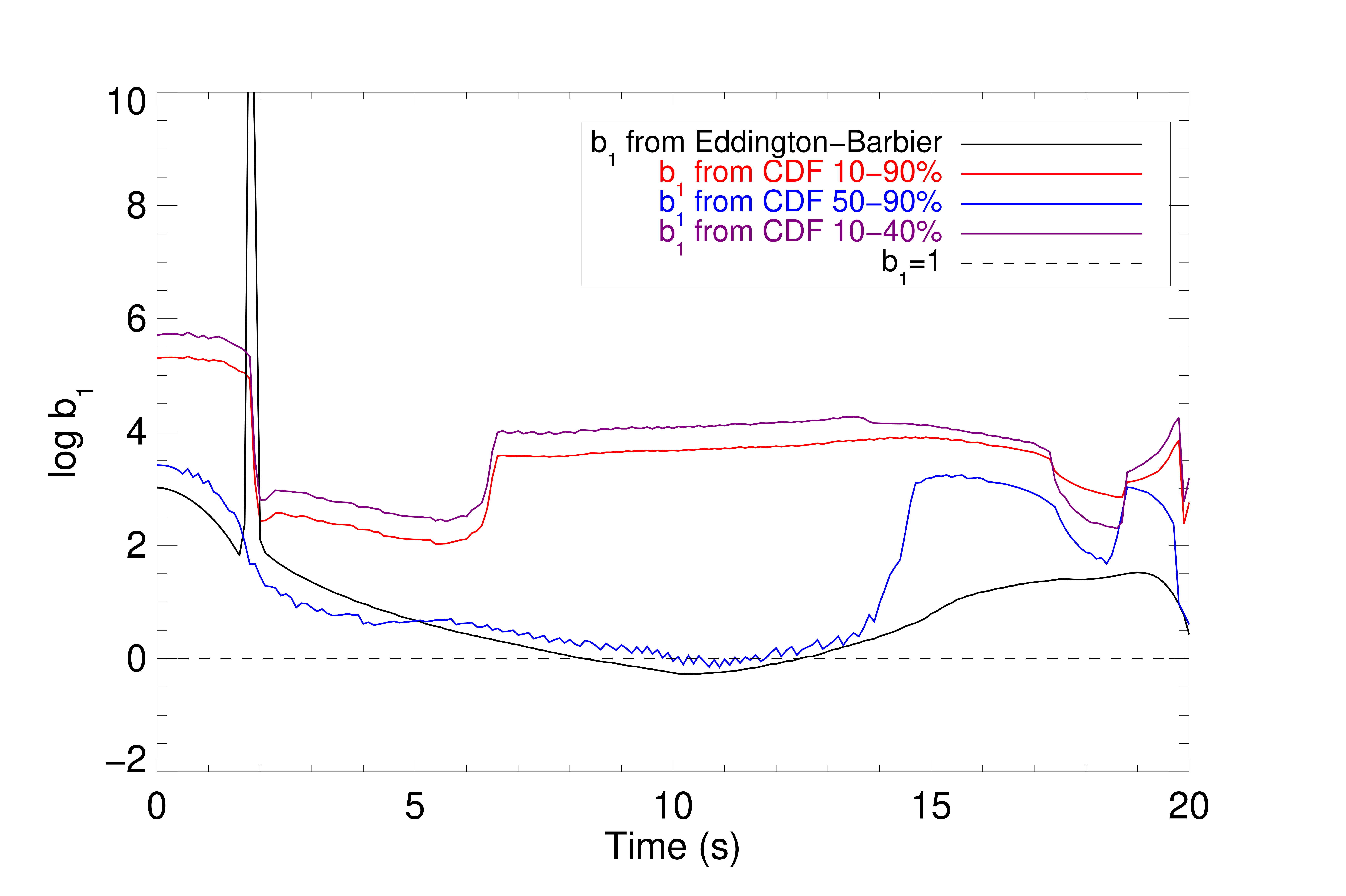}
\caption{Temporal evolution of $<b_{1,rad}>$ values for the $3F10$, $\delta=6$, $E_c = 20$~keV model. The $<b_{1,rad}>$ values are shown for where $\lambda = 900.3$~\AA\ emission originates,  considering $C_{cdf}$ = $[10$--$90\%,~50$--$90\%,~ 10$--$40\%]$ (red, blue, and purple, respectively).  Also shown is the result of fitting the spectra between $\lambda = 700$--$911$~\AA\ using the EB approximation (solid black line). The $b_1$=1 line is shown by the dashed black line for reference}
\label{b1_3e11_d6_cdf}
\end{figure}

Figure~\ref{b1_3e11_d6_cdf} shows $<b_{1,rad}>$ compared with $b_{1}$ obtained from our spectral fitting (solid black line). Formation height ranges corresponding to $[10$--$90\%,~50$--$90\%,~10$--$40\%]$ (red, blue, and purple lines, respectively) were considered. These ranges were selected by integrating the contribution function as a function of height and determining the ranges where the emission became optically thick or thin. The emission range $10$--$90\%$ considers both optically thick and thin components, $50$--$90\%$ considers emission between $\tau_{\lambda} \approx0.1$--$1$ (optically thick), and $10$--$40\%$ considers emission above the $\tau_{\lambda}<0.1$ layer (optically thin) at 900.3\AA. The $<b_{1,rad}>$ values from the optically thick layer are more consistent with those obtained from the spectra but deviate during the declining phase of the beam heating.

The $<b_{1,rad}>$ values determined over $10$--$90\%$ and $10$--$40\%$ of the NCDF are significantly larger than the values determined from the spectra. The former represents an assessment of what the $<b_{1,rad}>$ value when including both thick and thin LyC emitting regions. This suggests that the thin components (originating from the bubbles) have such large $<b_{1,rad}>$ values that they drag the overall average up considerably from the values in the chromosphere. The fact that $<b_{1,rad}>$ derived from the optically thick region is more consistent with $b_{1}$ derived from spectral fitting could be due to the optically thick emission dominating the emergent intensity.  %\gsk{GSK: regarding the statements about overdense bubbles, overdense compared to what? Do you mean denser than *actual* bubbles, then we should clarify, and mention that the Graham et al 2020 citation is due to them determining that the electron denisty in the condensations they produce is potentially larger than the observations would require.}  %There can be important optically thin contributions at coronal temperatures, so that the averaged $<b_{1,rad}>$ values are elevated as fluctuations in $b_1$ a few 100km above the main formation region are taken into account when the optically thin emission is considered. 

%The $10-90\%$ and $10-40\%$ $<b_{1,rad}>$ values are larger during the time of the spikes, while the $50-90\%$ values behave as expected. At these times, the $<b_{1,rad}>$ values are larger, when the optically thin layer is considered, as the region of minimal heating is taken into account in the averaging. As previously mentioned, $n_1$ in this region remains large and thus $<b_{1,rad}>$ too. This region is neglected in the $50–90\%$ averages, and thus $<b_{1,rad}>$ behaves as previously expected.

\begin{figure}[ht!]
\includegraphics[width=0.5\textwidth]{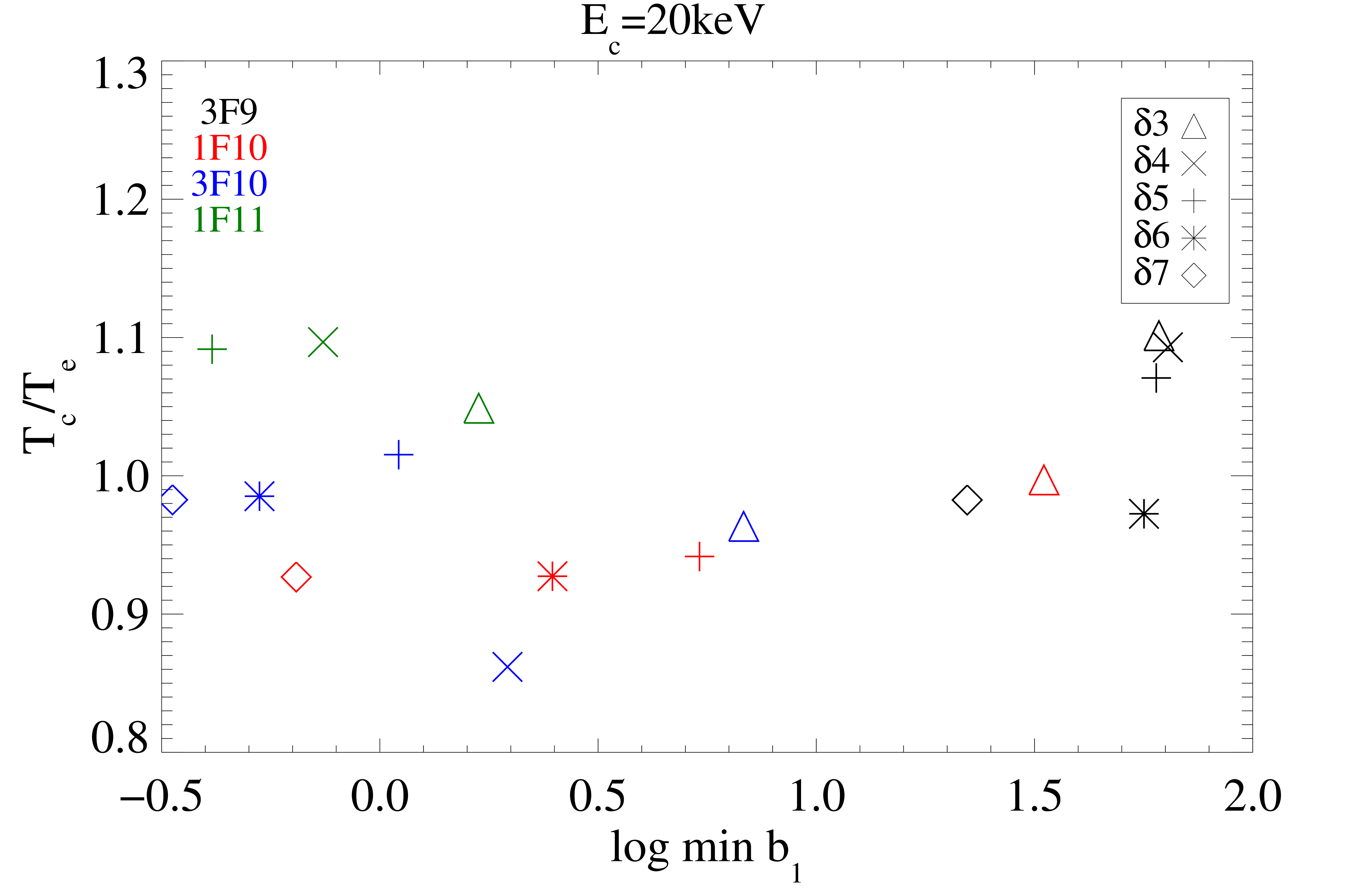}
\caption{The ratio of $T_c$ to the electron temperature ($T_e$), determined when $b_1$ is at a minimum, for the $3F9$, $1F10$, $3F10$, and $1F11$ models, with $\delta$=3--7, and $E_c$=20~keV where available. $T_e$ was determined from RADYN for emission between $50$--$90\%$ of the normalised contribution function at 900.3~\AA. $T_c$ and $b_1$ values were determined from fitting the spectra between $700.0$--$911.0$~\AA.}
\label{te_tc_vs_b1}
\end{figure}

%\begin{figure}[ht!]
%\includegraphics[width=0.5\textwidth]{FCRHOMA_HexHist_Tratio_vs_b1radyn_850A_010to095_750to911_time_CUT.png}
%\caption{The ratio of the electron temperature ($T_e$) to $T_c$ as a function of $b_1$ for all models, weighted by time. $T_e$ was determined from RADYN for emission between $10$--$95\%$ of the normalised contribution function at 850.0\AA. $T_c$ and $b_1$ values were determined from fitting the spectra between $750.0$--$911.0$\AA.}
%\label{te_tc_vs_b1}
%\end{figure}

%\begin{figure}[ht!]
%\includegraphics[width=0.5\textwidth]{Te_Tc_vs_b1_rate_multiple_new.eps}
%\caption{The ratio of the electron temperature ($T_e$) to $T_c$ as a function of the rate of change of $b_1$ for the $1F11$ models, with $E_c$=20keV, and $\delta$=3--5. $T_e$ was determined from RADYN for emission between $10$--$90\%$ of the normalised contribution function at 900.3\AA. $T_c$ and $b_1$ values were determined from fitting the spectra.}
%\label{te_tc_vs_rate_b1}
%\end{figure}

In a similar manner, the average electron temperature, $<T_{e}>$ from the LyC forming regions was calculated (replacing $b_{1,rad}$ for $T_{e}$ in Equation~\ref{eq:avvalue}).As $b_1$ tends towards unity, $T_c$ is expected to tend towards the electron temperature, $T_e$ \citep{Machado2018}. Figure~\ref{te_tc_vs_b1} shows the ratio of $T_c$ to $<T_e>$, determined when $b_1$ is at a minimum for the $3F9$, $1F10$, $3F10$, and $1F11$ models, with $\delta=5$--$7$, and $E_c=20$~keV outside of the times of the anomalous $b_1$ values. $<T_e>$ was determined over the optically thick layer (50--90\% of the $C_{cdf}$) as the $<b_{1,rad}>$ values over this emission range are in general agreement with the $b_1$ values obtained from the fits at these times. From Figure~\ref{te_tc_vs_b1}, the ratio of $T_c$ to $<T_e>$ is clustered around 1 when $b_1$ reaches a minimum. The minimum $b_{1}$ has a rather large range, extending from $b_{1}<1$ to $b_{1}\sim100$. That is, the ratio $T_c:<T_e>$ approaches unity at the minimum value of $b_{1}$ but $b_{1}$ itself does not necessarily have a value of unity, somewhat contrary to our expectations. \cite{Machado2018} present a similar finding where the 6 different X-class flares they observed had varying $b_1$ values ($b_{1}<1$ to $b_{1}\sim68$). However, as they used EVE Sun-as-a-star observations, they converted EVE spectral irradiances to specific intensities by assuming the flaring area. \cite{Machado2018} assumed a fixed flaring area of $10^{18}$~cm$^{-2}$, representing the middle of a rather large range of reported areas of X-class flares (see Section 4 of \citealt{Machado2018}). Varying this area would in effect shift the LyC specific intensity up or down, while keeping the spectral slope fixed. This would vary the $b_{1}$ value but not $T_{c}$. %However, the values of $T_c~/<T_e>$ do not converge closer to one for the models where the minimum value of $b_1$ is closer to unity (points going right to left in Figure~\ref{te_tc_vs_b1}). 

\begin{figure}[ht!]
\includegraphics[width=0.5\textwidth]{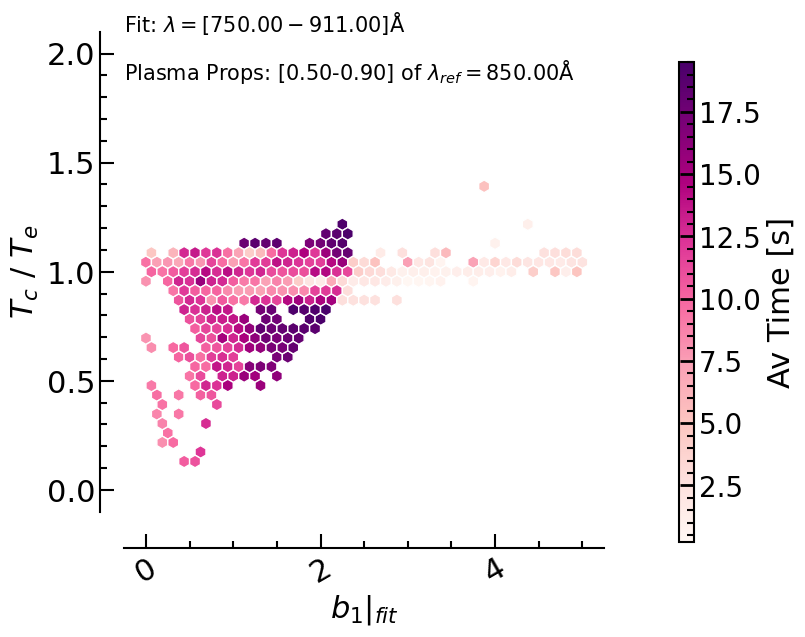}
\caption{The ratio of $T_c$ to electron temperature ($T_e$) as a function of $b_1$ for all models, weighted by time. $T_e$ was determined from RADYN for emission between $50$--$90\%$ of the normalised contribution function at 850.0\AA. $T_c$ and $b_1$ values were determined from fitting the spectra between $750.0$--$911.0$\AA.}
\label{Spread_Tc_Te}
\end{figure}

Outside of the times when $b_1$ is at a minimum, the ratio of $T_c$ to $T_e$ was found to have a large range of values. Figure~\ref{Spread_Tc_Te} shows the ratio of $T_c$ to $T_e$ as a function of $b_1$ for all models, weighted by time. As seen in Figure~\ref{Spread_Tc_Te}, even as $b_1$ approaches unity, the range of $T_c/T_e$ extends from $0.1$--$1.5$. This may be because $<T_e>$ is determined using the NCDF of the contribution function. As $<T_e>$ is determined over a given emission range, the heights considered will vary from model to model, particularly for the more energetic beams where the dynamics of the formation layers occur on shorter time scales (see Figures~\ref{contrib_3e10_d5} and \ref{contrib_1e11_d5}). Therefore, the value of $<T_e>$ is dependent on the height range considered.

%This is likely because $<T_e>$ is determined using the NCDF of the contribution function. As $<T_e>$ is determined over a given emission range, the heights considered will vary from model to model, particularly for the more energetic beams where the dynamics of the formation layers occur on shorter time scales (see Figures~\ref{contrib_3e10_d5} and \ref{contrib_1e11_d5}). Therefore, the value of $<T_e>$ is dependent on the height range considered.}

%mention that b1 is also

%However, we found that whilst the ratio of $<T_e>$ to $T_c$ tended towards one as $b_1$ approached unity, the ratio of one was not maintained when $b_1$ remained close to unity. Figure~\ref{te_tc_vs_rate_b1} shows the ratio of $<T_e>$ to $T_c$ as a function of the rate of change of $b_1$ (from the spectral fitting) for the $1F11$ models, with $E_c = 20$~keV, and $\delta = [3,4,5]$. For all three models, there is no correlation between the $<T_e>$ and $T_c$ when $b_1$ has a positive (i.e. $b_{1}$ increasing) or negligible rate of change. However, when there is a large negative gradient in $b_1$ (i.e. $b_{1}$ decreasing), $T_c$ tends towards $<T_e>$. $T_c$ is thus approximately equal to $<T_e>$ during the beam onset when the rate of change in $b_1$ is strongly negative. 

\begin{figure}
	\centering 
	%\hbox{
	%	\subfloat{\includegraphics[width = 0.5\textwidth, clip = true, trim = 0cm 0cm 0cm 0cm]{FCRHOMA_HexHist_Tg_vs_ne_radyn_850A_005to095.pdf}}
	%	}
    \hbox{
		\subfloat{\includegraphics[width = 0.5\textwidth, clip = true, trim = 0cm 0cm 0cm 0cm]{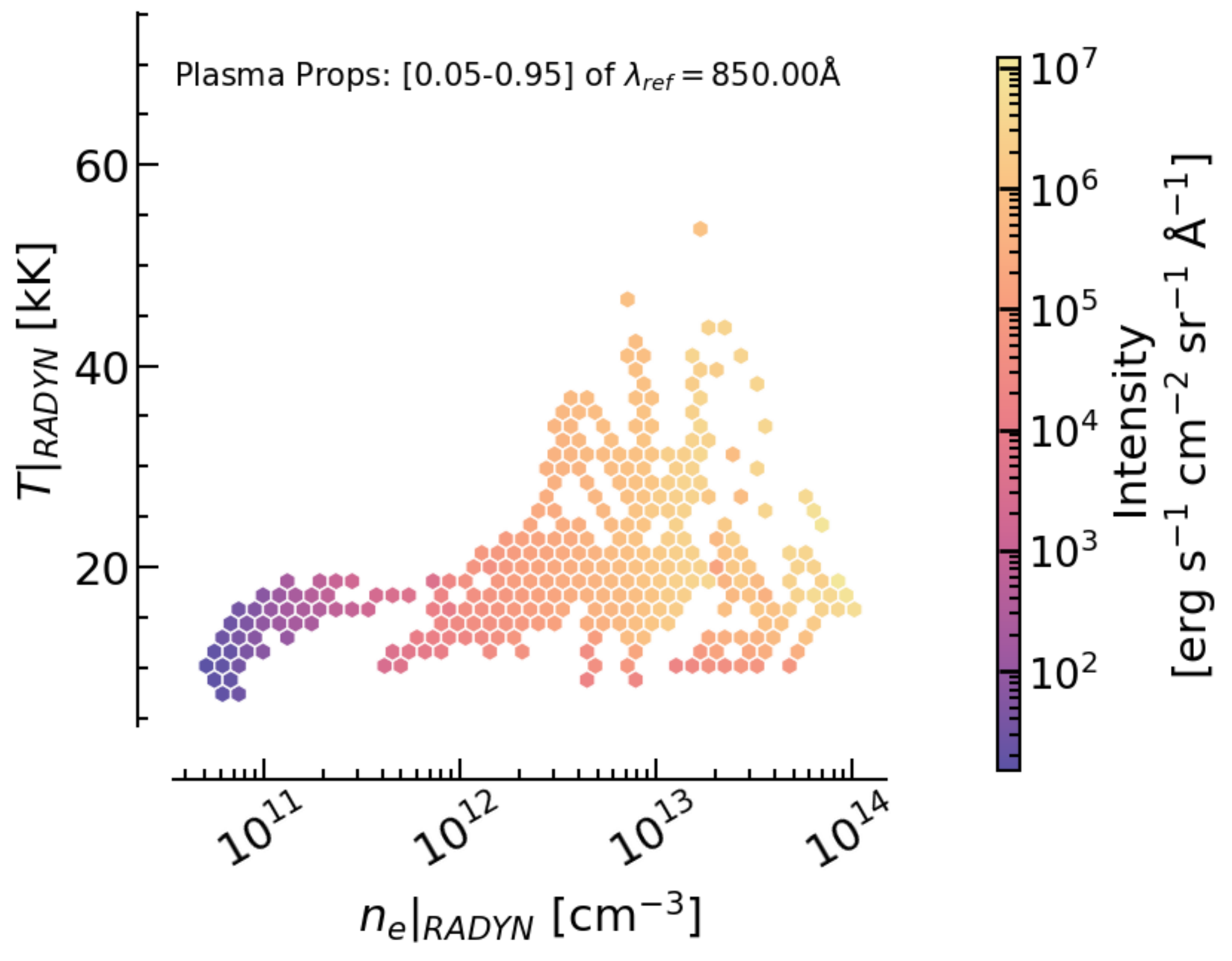}}
		}
	\caption{{Histogram of $<T_{e}>$ and $<n_{e}>$ in the formation region of $\lambda = 850$~\AA, weighted by the average intensity of the emission.}}
	\label{fig:formprop_hists}
\end{figure}

Finally, we determined $<T_{e}>$ and $<n_{e}>$ for $\lambda = 850$~\AA\ to illustrate the general range of plasma properties where LyC forms. Those are presented as a 2D histogram in Figure~\ref{fig:formprop_hists} in which properties from $\delta = [3,4,5]$, $E_{c} = [15, 20, 25]$~keV, $F_{peak} = [3F9, 1F10, 3F10, 1F11]$~erg~s$^{-1}$~cm$^{-2}$ were collated. Figure~\ref{fig:formprop_hists} shows that while there is some spread to high temperatures (due mostly to contributions from dense, optically thin bubbles) the bulk of the flare emission forms in the range $<T_{e}> = 10$--$30$~kK, with electron densities spanning $<n_{e}> = 10^{11-14}$~cm$^{-3}$. The data has been weighted by the average intensity of emission, and indicates that higher intensity is generally associated with higher $<n_{e}>$, though temperature also plays a role.

%The top panel shows that while there is some spread to high temperatures (due mostly to contributions from dense, optically thin, bubbles) the bulk of the flare emission forms in the range $<T_{e}> = [10-30]$~kK, with electron densities spanning $<n_{e}> = 10^{11-14}$~cm$^{-3}$. The bottom panel is weighted by the average intensity of emission, and indicates that higher intensity is generally associated with higher $<n_{e}>$, though temperature plays a role also.

%up to here;;;;;;;;;;;;;;;;;

\section{Discussion}\label{DISCUSSION} %corrected up to here so far

\subsection{Percentage of energy radiated away by the LyC}

\begin{figure}[ht!]
\includegraphics[width=0.5\textwidth]{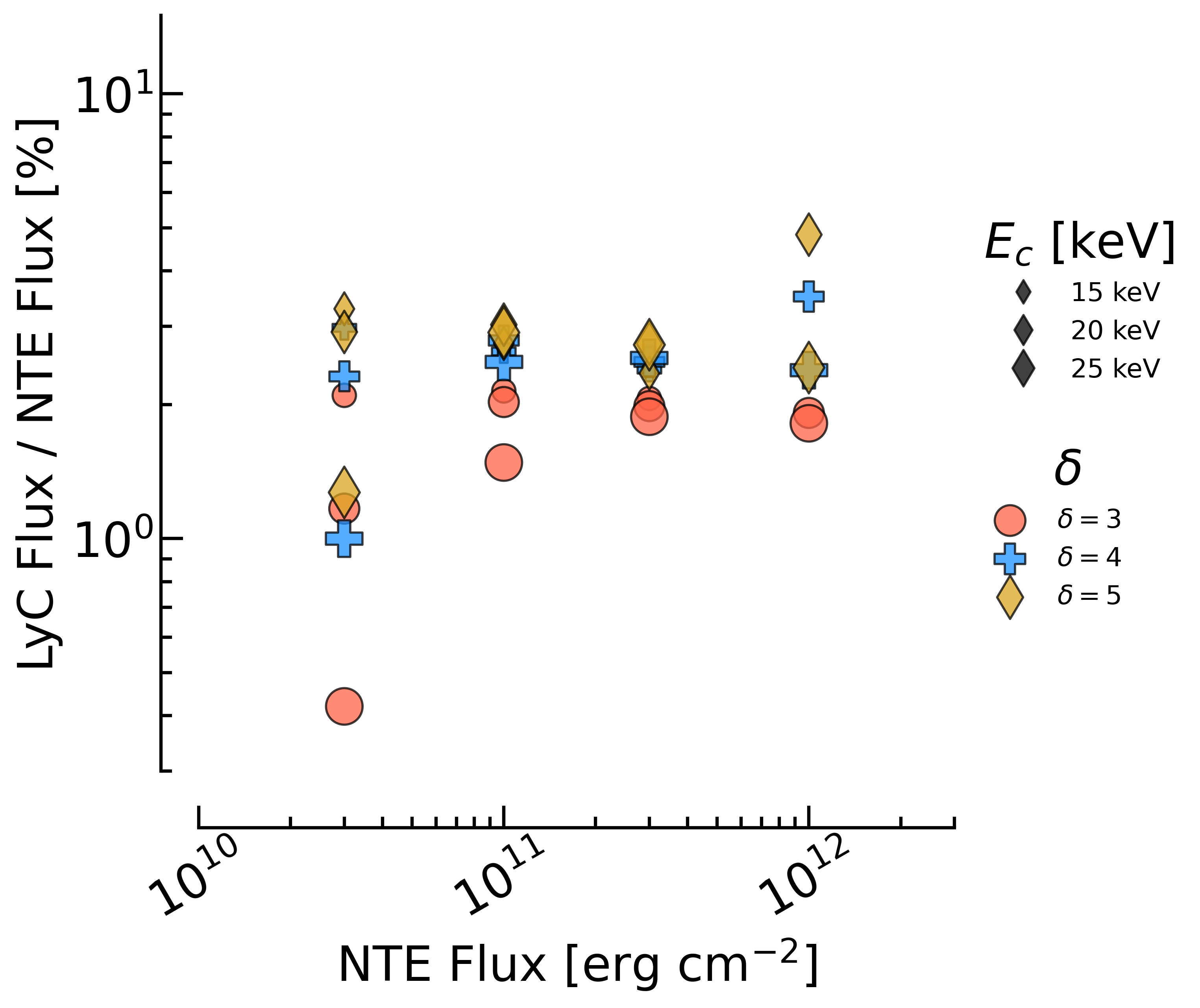}
\caption{The ratio of energy radiated by LyC to the energy injected by non-thermal electrons (NTE), as a function of the non-thermal electron energy. Circles represent $\delta = 3$, plus symbol $\delta = 4$, and diamonds $\delta = 5$. The symbols increase in size with increasing $E_{c}$.}
\label{fig:lycenergyratio}
\end{figure}

To determine the percentage of energy radiated by the LyC compared to the total energy injected via non-thermal electrons, the LyC lightcurves in Figure~\ref{LyC_lightcurves} were integrated over time and divided by the total non-thermal electron energy. We found the LyC radiated away between $1$--$3\%$ of the total non-thermal electron energy injected. The $3F9$ models radiated away around $1\%$ of the total injected energy through the LyC, whereas the $3F10$ models radiated around $3\%$. The overall distribution from a number of simualtions is shown in Figure~\ref{fig:lycenergyratio} %This is because beams of larger non-thermal electron fluxes will ionise a greater number of hydrogen atoms. Therefore, more recombinations occur. 

\cite{Milligan2014} provided a study of the global energy budget in a strong flare, comparing the energy radiated in the lower solar atmosphere at optical, UV, and EUV wavelengths to the energy injected via non-thermal electrons. They found the LyC radiated away approximately $1\%$ of the total non-thermal electron energy. This is in agreement with our findings that predict LyC to radiate away a few percent of the total non-thermal electron energy.

\begin{figure}[ht!]
 \includegraphics[width=0.5\textwidth]{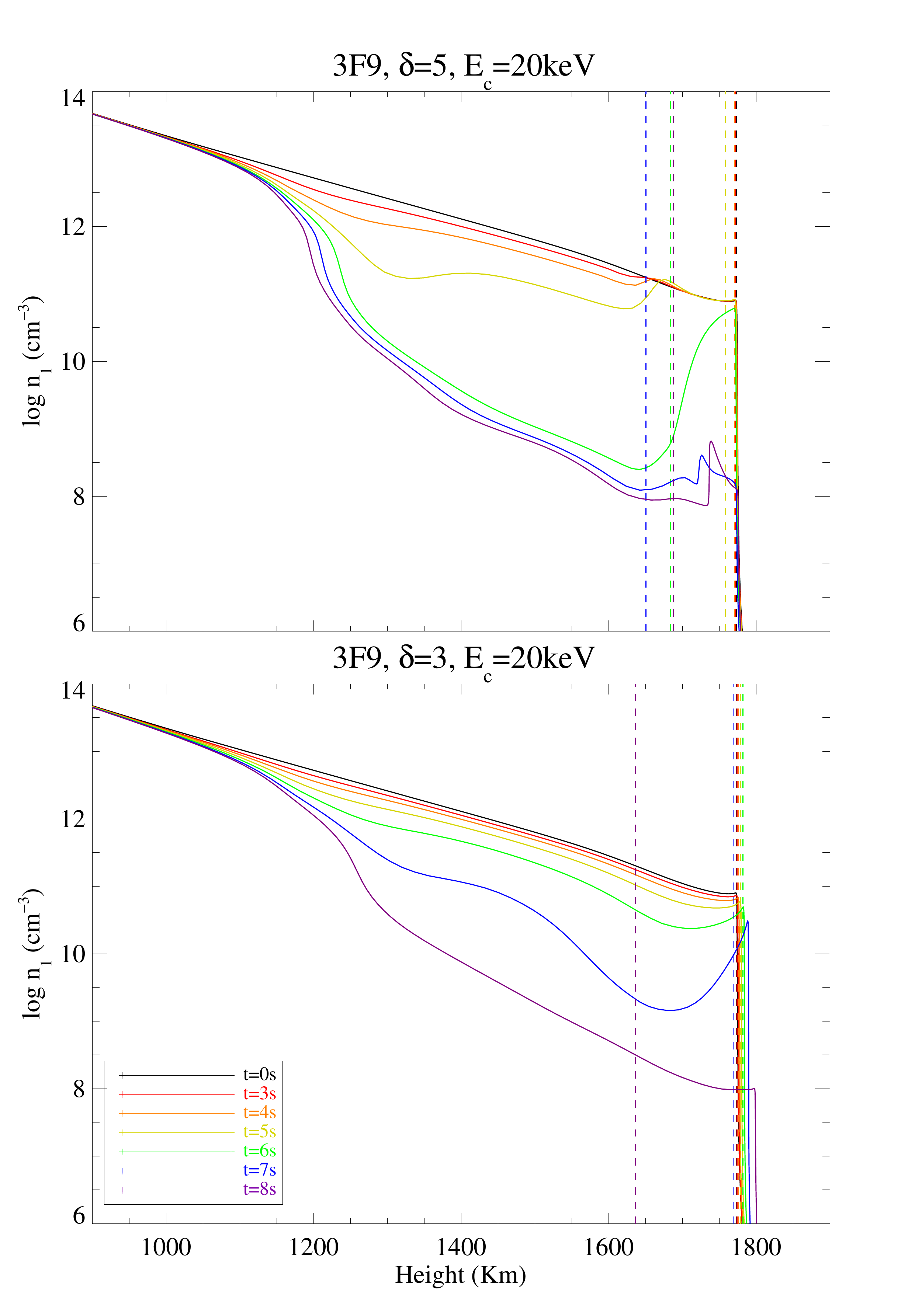}
 \caption{The temporal evolution of the ground state of hydrogen level population for the $3F9$ models, with $E_c = 20$~keV, and $\delta = [3,5]$. The height of the peak of the LyC contribution function is shown by the vertical dashed lines at each time step.}
 \label{n1_plot}
 \end{figure}

\subsection{Spikes in the Departure Coefficient and color Temperature values}\label{SPIKES}
%\begin{figure}
%	\centering 
%	\hbox{
%		\subfloat{\includegraphics[width = 0.475\textwidth, clip = true, trim = 0cm 0cm 0cm 0cm]{LyC_HPops_3F9d3Ec20_varytimes}}
%		}
%    \hbox{
%		\subfloat{\includegraphics[width = 0.475\textwidth, clip = true, trim = 0cm 0cm 0cm 0cm]{LyC_HPops_3F9d5Ec20_varytimes}}
%		}
%	\caption{{The temporal evolution of the ground state of hydrogen level population for the $3F9$ models, with $E_c = 20$~keV, and $\delta = [3,5]$.}}
%	\label{n1_plot}
%\end{figure}
We mentioned several times previously the appearance of strong spikes in $b_1$ and $T_c$, caused by a flattening towards the head of the LyC, as seen, for example, at $t = 5.7$~s in the left-hand panel of Figure~\ref{Example_EB_fit}. The Eddington-Barbier approximation is clearly no longer valid during these times. 

This phenomenon can be understood from Figure~\ref{n1_plot}, which shows the temporal evolution of the ground state of hydrogen level population for the $3F9$ models, with $E_c = 20~$keV, and $\delta = [3,5]$. The height of the peak of the LyC contribution function is shown by the vertical dashed lines at each time step. At $t = 0$~s, $n_1$ is large in the chromosphere and decreases across the transition region for both models. As the beam heating begins, $n_1$ decreases within the chromosphere due to excitation and ionization following the temperature increase and non-thermal collisions. However, there is a small region of plasma between the beam heating region and the transition region that is only minimally heated by the non-thermal electrons, resulting in $n_1$ remaining large compared to the adjacent plasma. At later times, the optically thick layer shifts much deeper into the chromosphere. The photons emitted from the optically thick layer subsequently get absorbed by this region of plasma, resulting in the flattening of the head of the LyC, and the delayed enhancement of the LyC lightcurves. For the $\delta = 3$ model, the optically thick layer of the LyC forms below the region of minimal heating for a few seconds until the region dissipates, resulting in the extended increase in $b_1$ and $T_c$ observed for harder beams (see $\delta$=3--4 curves in Figure~\ref{b1_Tc_fits}). Whereas for softer beams (see $\delta \geq$5 curves in Figure~\ref{b1_Tc_fits}), the optically thick layer only forms below the region of minimal heating for a shorter duration, resulting in the sudden steep spikes in $b_1$ and $T_c$.

%This region lasts for a few seconds in the $\delta = 3$ simulation, resulting in the extended spike in $b_1$ and $T_c$ observed for harder beams ($\delta$=3-4), whereas for softer beams ($\delta \geq$5), this region persists only for a short duration. 

\begin{figure}[ht!]
\includegraphics[width=0.5\textwidth]{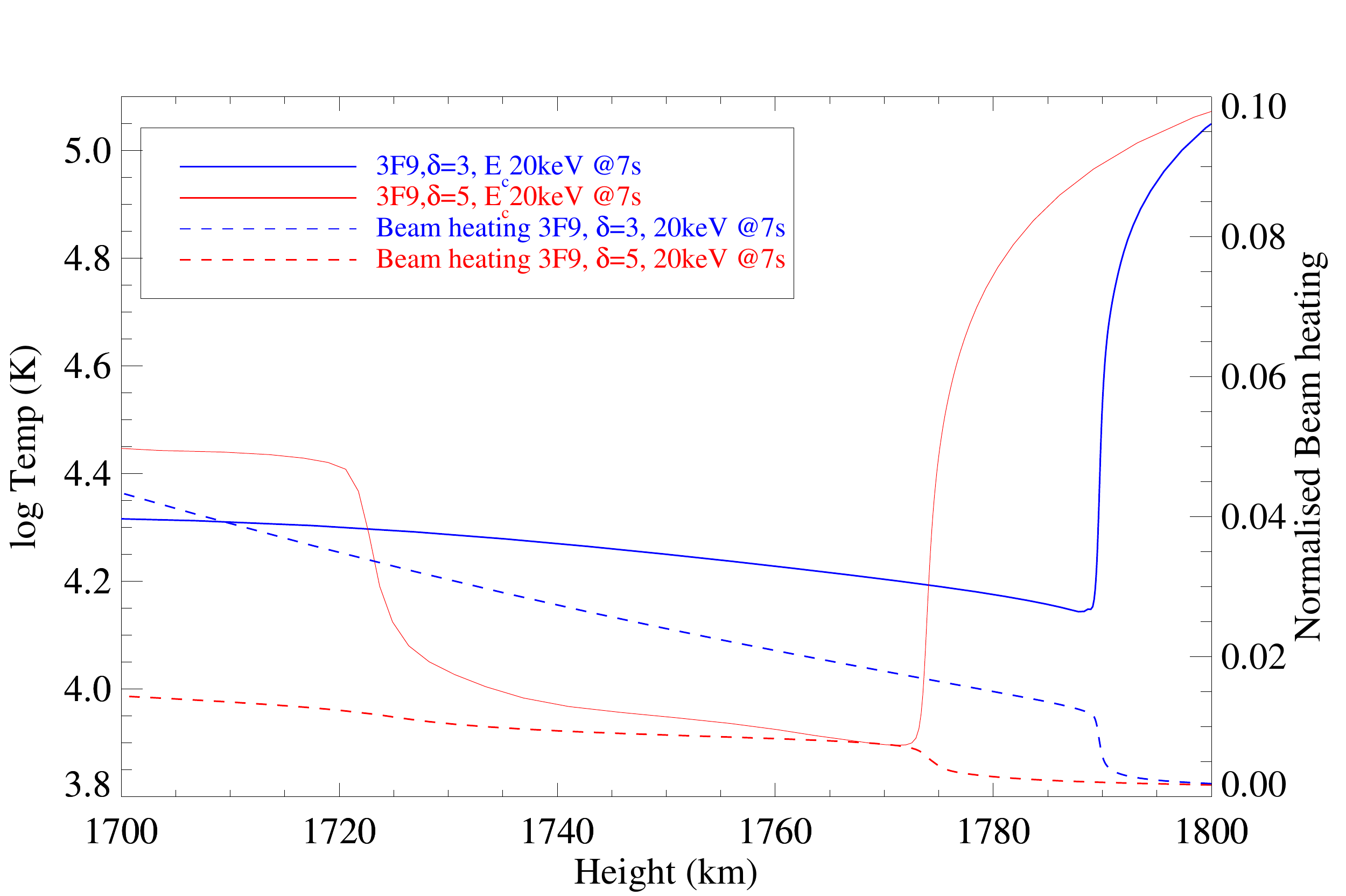}
\caption{Temperature profiles for the $3F9$ models, with $E_c=20$~keV, and $\delta=3$ and 5 at t=7~s. The normalised beam heating is shown by the dashed lines for both models.}
\label{Temp_profiles_spikes}
\end{figure}

In Figure~\ref{n1_plot}, the $n_1$ values for the harder beam ($\delta$=3) are reduced more uniformly and at higher altitudes within the chromosphere despite harder beams being composed of a greater number of high energy, deeply penetrating, electrons. This can be understood from Figure~\ref{Temp_profiles_spikes}, which shows the temperature and beam heating profile for the $3F9$ models, with Ec=20keV, and $\delta$=3 (harder beam; blue lines) and 5 (softer beam; red lines). The region of minimal heating is thicker for softer beams ($z\approx$1720--1770~km). The $\delta$=3 beam is harder and therefore has a greater number of high-energy electrons, which results in a faster atmospheric response and evaporation of chromospheric plasma. As the evaporation front propagates the mass density of the upper atmosphere is increased by approximately a few orders of magnitude. The increased column depth means that the higher energy electrons in the $\delta$=3 beam are thermalised higher up in the atmosphere at later times in the simulation despite the $\delta$=5 beam being composed of more low energy electrons, as can be seen by the blue and red dashed lines in Figure~\ref{Temp_profiles_spikes}. This results in the narrower region of minimal heating ($z\approx$1790~km) seen for the harder $\delta$=3 case. The thickness of this region also contributes to the varying profiles during the sudden increases (spikes) in $b_1$ and $T_c$.

\subsection{Wavelength dependency of $b_1$ and $T_c$}\label{Two_gradients}

\begin{figure}[ht!]
\includegraphics[width=0.5\textwidth]{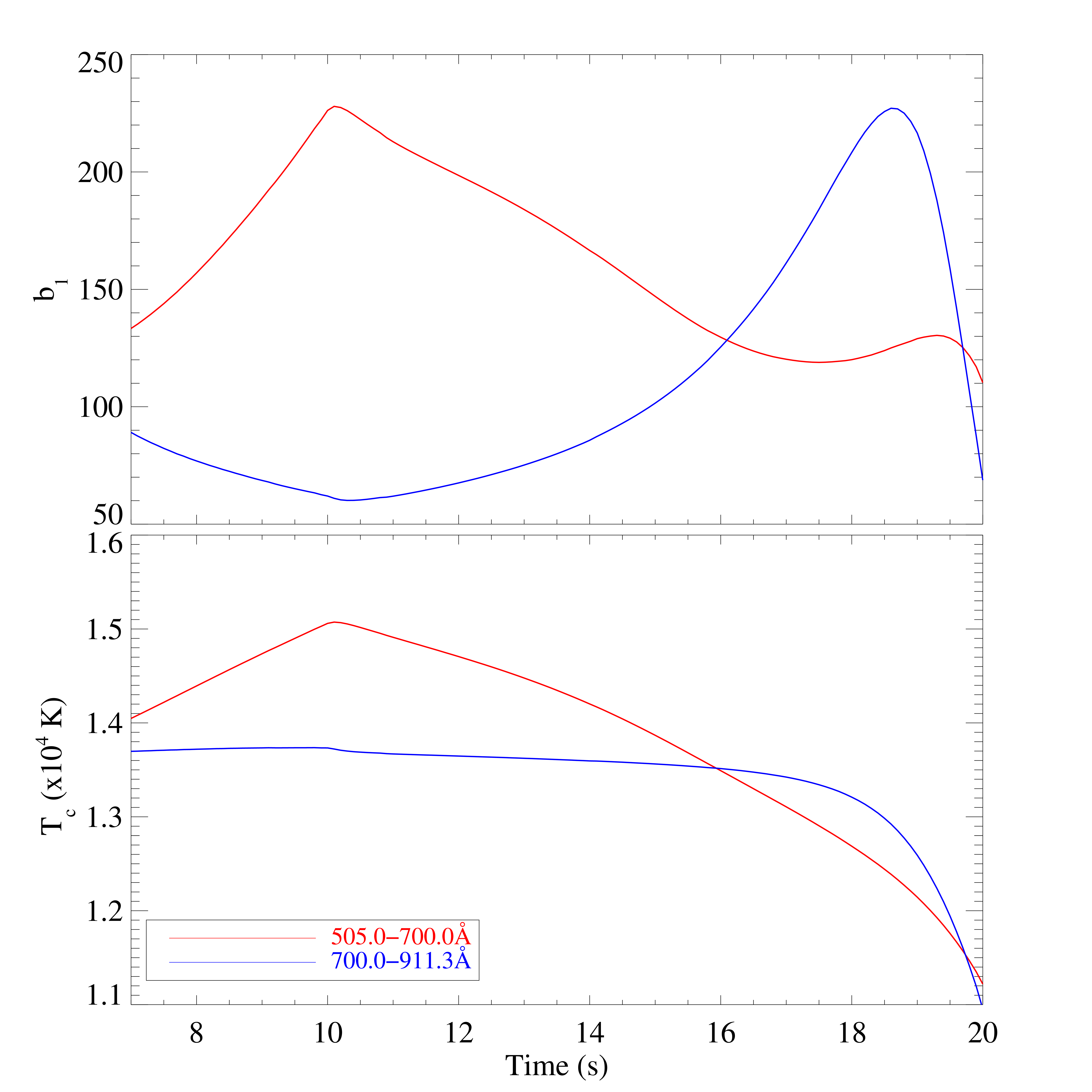}
\caption{Evolution of $b_1$ and $T_c$ for the $3F9$, $\delta = 5$, $E_c = 20$~keV model. Equations~\ref{EQN:E-B} and \ref{EQN:Planck} have been applied between $\lambda = [505-700$]~\AA\ and $\lambda = [700-911$]~\AA. The values are only shown between $t = [7 - 30]$~s to omit times where we know that spectral fitting produced poor results.}
\label{Chi_square_2_fits}
\end{figure}

\cite{Machado1978} hypothesised from observational evidence that a higher-lying region contributes optically thin component to the LyC during flares alongside the bulk of the chromospheric emission. Evidence for this optically thin layer comes from a steepening in the LyC spectrum's gradient away from the continuum head, resulting in increased $T_c$ values determined at shorter wavelengths. In the right-hand panel of Figure~\ref{Example_EB_fit}, there is a local peak in $\epsilon$ at t=10s. Shorter wavelengths, between $\lambda\sim505.0-700.0$\AA, are poorly fit compared to longer wavelengths. This spectrum has two distinct gradients, one at shorter wavelengths, $\lambda\leq 700$~\AA, and the other at longer wavelengths, $\lambda\geq 700$~\AA, in agreement with \citet{Machado1978}, \citet{Machado2018}, and \cite{Druett2019}. 

Figure~\ref{Chi_square_2_fits} shows the $b_1$ and $T_c$ values for the $3F9$, $\delta = 5$, $E_c = 20$~keV model, where the EB approximation has been applied between $\lambda = 505-700$~\AA\ and $\lambda = 700-911$~\AA. Generally, the $T_c$ values determined at shorter wavelengths were a few thousand Kelvin hotter than at longer wavelengths, in agreement with the literature \citep{Machado1978,Machado2018}.  The $b_1$ values are also generally larger at shorter wavelengths. This is due to the optically thin components of the LyC that enhance the spectrum at shorter wavelengths. The number of optically thin components that form depends on the type of evaporation observed. Gentle evaporation resulted in one upwardly propagating optically thin layer forming. Whereas, explosive evaporation resulted in two or three optically thin components corresponding to evaporation and condensation fronts. The upward propagating optically thin components of the LyC in Figure~\ref{contrib_1e11_d5} are due to bubbles of chromospheric material travelling immediately ahead of the evaporation front. 

\citet{Reid2020} state that these bubbles are a source of optically thin  Ca II 8542\AA\ line emission and do not always emit strongly in the H$\alpha$ line, while \citet{Brown2018} found solar bubbles to be among the dominant sources of Lyman alpha line emission. Such propagating high-density features could be indirectly detected via LyC observations due to the effect they have on the spectral shape, manifesting as an increase in both $T_c$ and $b_1$ at shorter wavelengths. \citet{Machado2018} only observed an increase in $T_c$ at shorter wavelengths for one of the six events that they analysed. This may be due to the fact that the other five flares did not provide the correct conditions for a solar bubble to form, or at the time of observation the bubbles had dissipated.

\section{Conclusions}\label{CONCLUSION}

Using the F-CHROMA grid of RADYN models, we have shown that the LyC is greatly enhanced during solar flares. The LyC spectral response is highly sensitive to the flux of the non-thermal electron beam, but is less dependent upon the spectral index or low-energy cutoff. LyC was found to radiate away between 1–3\% of the total non-thermal electron energy injected. Increases in solar irradiance associated with solar flares are known to drive dynamic and compositional changes in Earth’s ionosphere, which can have adverse implications for modern technology on which society has become dependent. The $850$--$1027$\AA\ range, in particular, is absorbed at an altitude of around $105$--$120$km in the ionosphere (E-layer), where it drives the partial dissociation of molecular oxygen \citep{LyC_space_weather}. This part of the spectrum is dominated by LyC, along with higher-order Lyman emission lines.  

Both optically thin and thick layers of the LyC were found to form during solar flares, in agreement with the literature \citep{Machado1978,Machado2018,Druett2019}. The optically thick layer is formed in NLTE in the QS and forms at the top of the chromosphere. During solar flares, this layer shifts deeper into the solar chromosphere due to the evaporation of the upper chromosphere, forming near the peak beam heating region, with $<T_e> \approx10$--$40$~kK. It also forms closer  to LTE conditions. The optically thin components of the LyC formed due to chromospheric evaporation, with the number of optically thin components forming being dependent on the type of evaporation observed, gentle or explosive. These optically thin components cause an enhancement in intensities away from the LyC head, resulting in increased $b_1$ and $T_c$ values determined at shorter wavelengths. Fitting with the EB relation we find that $T_{c} \approx 10$--$16$~kK, with $b_{1}$ at times dropping to $b_{1} \approx \mathrm{few} \times 10$, but which have a large scatter. 

Our results suggest that the LyC spectral response is indicative of a chromospheric temperature and density enhancement largely probing the chromosphere. Gradients in the derived $T_{c}$ as a function of wavelength can also indicate the presence of regions of propagating dense, cool material in the upper atmosphere. The number of optically thin layers formed was found to be greater for stronger solar flares, and the LyC contribution functions presented show that the optically thin layers form and dissipate over a shorter period for stronger flares. Model-data discrepancies (e.g. the large scatter of $b_{1}$ values that don't always approach unity, contrary to \citealt{Machado2018}), could result from an exaggerated optically thin component to the LyC forming in the dense bubbles, if those bubbles are denser than in actual flares. For example, \cite{2020ApJ...895....6G} modelled the ratio of the intensity of redshifted `satellite' components of the \ion{Fe}{2} line to the intensity of the stationary component, noting that it was larger than the observed flare they were simulating. This could be due to the density in the modelled condensation (from which the satellite component originated) being larger than the condensation produced during the observed flare. Our chromospheric bubbles may be similarly over-dense.

EVE currently provides LyC observations with the greatest coverage.However, these are Sun-as-a-star observations with a cadence of 60~s (observations prior to 2014 had a 10~s cadence, but instrument degradation has forced a longer exposure time). Therefore, EVE LyC flare observations will observe a range of flaring loops at various heating or cooling stages. Thus, some of the dynamic features may become smeared temporally, especially for stronger flares. As the breakdown in the EB approximation and delayed enhancement in the LyC lightcurves also occur over second to sub-second timescales, it is unlikely that EVE can observe these phenomena. Additionally, as EVE provides Sun-as-a-star observations, EVE spectral irradiances can be converted to specific intensities by assuming the flaring area as discussed in Section~\ref{model_fit_comparison}. Varying this area shifts the LyC specific intensity up or down, while keeping the spectral slope fixed. This would vary the $b_{1}$ value but not $T_{c}$.

%EVE currently provides LyC observations with the greatest coverage. However, EVE has a cadence of 60s (observations prior to 2014 had 10~s cadence, but instrument degradation has forced a longer exposure time). Therefore, some of the dynamic features may become smeared temporally, especially for stronger flares. As the breakdown in the EB approximation \gsk{and delayed enhancement in the LyC lightcurves also occur over second to sub-second timescales, it is unlikely that EVE can observe these phenomena. As EVE provides Sun-as-a-star observations with a cadence of 60~s, EVE LyC flare observations will observe a range of flare loops at various heating or cooling stages. Therefore, the more global flare LyC properties will}

SPICE on board the Solar Orbiter mission that was launched in 2020, provides EUV coverage in the wavelength ranges of $\lambda = 704$--$790$~\AA\ and $\lambda = 973$--$1049$~\AA. This provides partial coverage of LyC. It remains to be seen to what extent it is possible to extract continuum intensities, which will depend in part on how well we can resolve spectral lines with that passband, but SPICE observations may be used to determine $b_1$ and $T_c$ values below the head the LyC. Unlike SDO/EVE, these observations would have spatial resolution and obtain higher cadences. However, as SPICE only provides partial coverage of LyC, the derived $b_1$ and $T_c$ values may be elevated due to the presence of the optically thin LyC layers, enhancing the LyC spectrum at shorter wavelengths. To determine if this is the case, $b_1$ and $T_c$ values should be derived and compared from SPICE and EVE spectra over multiple wavelength ranges for any flares both instruments capture. Our analysis paves the way for an interpretation of solar flare LyC observations taken by current and future missions.

\begin{acknowledgments}
We thank the anonymous referee for a careful review of our manuscript, and for useful comments and suggestions. SAM would like to thank the Science and Technology Facilities Council (UK) for the award of a PGR studentship. AJM acknowledges funding from the Science Technology Funding Council (STFC) Grant Code ST/T506369/1. ROM and GSK would like to acknowledge support from NASA Heliophysics Supporting Research grant NNH19ZDA001N and ROM thanks the Science and Technologies Facilities Council (UK) for the award of an Ernest Rutherford Fellowship (ST/N004981/2). PJAS acknowledges support from the Fundo de Pesquisa Mackenzie (MackPesquisa), CNPq (contract 307612/2019-8), and FAPESP, the
Sao Paulo Research Foundation (contract 2013/24155-3). GSK thanks Dr. Adam Kowalski for sharing some analysis software that was adapted to perform this work. 
\end{acknowledgments}

%% For this sample we use BibTeX plus aasjournals.bst to generate the
%% the bibliography. The sample631.bib file was populated from ADS. To
%% get the citations to show in the compiled file do the following:
%%
%% pdflatex sample631.tex
%% bibtext sample631
%% pdflatex sample631.tex
%% pdflatex sample631.tex

\bibliography{LyC_Bib.bib}{}
\bibliographystyle{aasjournal}

\end{document}